\documentclass[10pt]{iopart}

\usepackage{iopams} 
\usepackage{graphicx}
\usepackage{hyperref}
\usepackage{url}
\usepackage{euclid}
\RequirePackage{siunitx}
\RequirePackage{xspace}

\usepackage[square]{natbib}
\begin{document}

\title[A proposal for relative in-flight flux self-calibrations for spectro-photometric surveys]{A proposal for relative in-flight flux self-calibrations for spectro-photometric surveys}

\author{S~Davini$^1$, I~Risso$^{1,2}$,
 M~Scodeggio$^3$, L~Paganin$^{1,2}$, S~Caprioli$^1$, M~Bonici$^{1,2}$, A~Caminata$^{1}$, S~Di~Domizio$^{1,2}$, G~Testera$^1$, S~Tosi$^{1,2}$, B~Valerio$^{1,2}$, M~Fumana$^3$, P~Franzetti$^3$.}
\address{$^1$ INFN Sezione di Genova, Via Dodecaneso 33, 16146 Genova, Italy}
\address{$^2$ Dipartimento di Fisica, Università degli Studi di Genova, Via Dodecaneso 33, 16146 Genova, Italy}
\address{$^3$ INAF-IASF Milano, Via Alfonso Corti 12, I-20133 Milano, Italy}
\ead{stefano.davini@ge.infn.it}

\begin{abstract}
We present a method for the in-flight relative flux self-calibration of a spectro-photometer instrument, general enough to be applied to any upcoming galaxy survey on satellite.
The instrument response function, that accounts for a smooth continuous variation due to telescope optics, on top of a discontinuous effect due to the segmentation of the detector, is inferred with a $\chi^2$ statistics.
The method provides unbiased inference of the sources count rates and of the reconstructed relative response function, in the limit of high count rates.
We simulate a simplified sequence of observations following  a spatial random pattern and realistic distributions of sources and count rates, with the purpose of quantifying the relative importance of the number of sources and exposures for correctly reconstructing the instrument response. We present a validation of the method, with the definition of figures of merit to quantify the expected performance, in plausible scenarios. 
\end{abstract}
\ioptwocol

\def\newblock{\ }
\newcommand{\fov}{0.5\,$\mbox{deg}^2$}
\newcommand{\expvisnisps}{565\,\mbox{s}} %
\newcommand{\mab}{m_{\rm AB}}
\newcommand{\mablow}{17}
\newcommand{\mabhigh}{12}
\newcommand{\code}[1]{\texttt{#1}}
\newcommand{\fpxy}{(x_i, y_i)}
\newcommand{\fpxyk}{(x_{k(i)}, y_{k(i)})}
\newcommand{\meanik}{\mu_{k(i)}}
\newcommand{\cnt}{c_{k(i)}}
\newcommand{\rate}{r_{k}}
\newcommand{\drate}{\delta \rate}
\newcommand{\sig}{\sigma^2_{k(i)}}
\newcommand{\noise}{n_i}
\newcommand{\resp}{f(x,y)}
\newcommand{\qvec}{\vec{q}}
\newcommand{\dqvec}{\delta \vec{q}}
\newcommand{\gvec}{\vec{g}}
\newcommand{\dgvec}{\delta \vec{g}}
\newcommand{\wvec}{\vec{w}}
\newcommand{\wvect}{\vec{w^T}}
\newcommand{\Hqr}{H_{\ell k}^{\rm(qr)}}
\newcommand{\covqqvec}{\tens{C}^{\rm(qq)}}
\newcommand{\covqgvec}{\vec{c}^{\rm(qg)}}
\newcommand{\reco}{\hat f(x,y \, | \qvec)}
\newcommand{\rreco}{\hat f(x,y \, | \qvec, \gvec)}
\newcommand{\recofp}{\hat f(x_i, y_i \, | \qvec)}
\newcommand{\rrecofp}{\hat f(x_i, y_i \, | \qvec, \gvec)}
\newcommand{\recofpk}{\hat f(x_{k(i)}, y_{k(i)} \, | \qvec)}
\newcommand{\rrecofpk}{\hat f(x_{k(i)}, y_{k(i)} \, | \qvec, \gvec)}
\newcommand{\Thetas}{\Theta_s}
\newcommand{\Thetasxy}{\Theta_s (x, y)}
\newcommand{\Thetaspxy}{\Theta_{s^\prime} (x, y)}
\newcommand{\Thetasxyk}{\Theta_s \fpxyk}
\newcommand{\ki}{\chi^2}
\newcommand{\ql}{q_\ell}
\newcommand{\dql}{\delta \ql}
\newcommand{\basis}{w_\ell}
\newcommand{\wlik}{w_{\ell(i,k)}}
\newcommand{\wmik}{w_{m(i,k)}}
\newcommand{\suml}{\sum_\ell}
\newcommand{\gs}{g_s}
\newcommand{\gsik}{g_{s(i,k)}}
\newcommand{\dgs}{\delta \gs}
\newcommand{\kisq}{\chi^2}
\newcommand{\kisqe}{\hat \chi^2}
\newcommand{\ndf}{N_{\rm dof}}
\newcommand{\nk}{n_k}
\newcommand{\nks}{n_k^{(s)}}
\newcommand{\sumk}{\sum_k}
\newcommand{\sumik}{\sum_i^{\nk}}
\newcommand{\sumiks}{\sum_i^{\nks}}
\newcommand{\Hlm}{H_{\ell m}^{\rm(q)}}
\newcommand{\Dvec}{\vec{\Delta}}
\newcommand{\dkisqdr}{\frac{\partial \kisqe}{\partial \rate}}
\newcommand{\dkisqdq}{\frac{\partial \kisqe}{\partial \ql}}
\newcommand{\dkisqdg}{\frac{\partial \kisqe}{\partial \gs}}

\section{Introduction}
Reliable determinations of fluxes and distances are of paramount importance for all large scale galaxy surveys.
The same source in the sky (e.g. a star, or a galaxy), observed in different positions on the focal plane, is typically recorded with different count rates.
Besides the statistical fluctuations of signal counts and background noise, the detected count rates of the same source will also differ because of the instrument response function dependency on the focal plane position; the dependency of the response function is due both to the optical distortions produced by the telescope optics and large-scale variations in the detector gain.

The non-ideal instrument response provides a systematic distortion of the source count rates, propagated as systematic errors on fluxes and magnitudes. In order to compensate for this systematic effect and provide accurate catalogues, the response function on the focal plane must be accurately determined. 

Several missions are foreseen in the next few years to build a three-dimensional map of the Universe by measuring positions of distant astrophysical sources and their fluxes and spectra. The European Space Agency will launch the Euclid satellite in 2022~\citep{redbook}. Euclid aims at providing a weak-lensing and spectro-photometric survey of a $15\,000$~deg$^2$ area of the extra-galactic sky, up to redshifts of about 2, and map the geometry of the Universe and the growth of structures~\citep{Amendola2018}. NASA is developing the Nancy Grace Roman Space Telescope (formerly known as WFIRST), whose launch is currently scheduled for 2025~\citep{roman}.
The Roman Space Telescope will use baryon acoustic oscillations, observations of distant supernovae, and weak gravitational lensing to probe dark energy.

The selection of a reliable galaxy sample in a survey  heavily relies on an accurate flux calibration. 
Contamination of the sample selection is caused by
all the effects which systematically vary the magnitude limit of the sample across the focal plane.
This contamination may bias the cosmological inference on the data, for example by injecting spurious signals in the galaxy clustering power spectrum within baryon acoustic oscillation measurements~\citep{Shafer2015}.
To mitigate this effect, an accurate flux calibration is needed.

In this work we specifically focus on the \emph{relative in-flight self-calibration} of the instrument response of a generic spectro-photometer as part of a satellite payload. In space missions, in light of the tight observation schedules, having an optimized and automated procedure to derive the response function is of crucial importance. 

In-flight calibration techniques exploit multiple measurements of bright sources recorded at different focal plane positions, obtained with partially overlapping exposures.
The relative instrument response can be inferred by the requirement that each source is reconstructed with statistically consistent count rates, within the whole focal plane.
The self-calibration method can accurately constrain the instrument response (and the source rates) when enough sources and exposures are provided.
The \emph{relative} flux calibration is determined up to a multiplicative global scale factor (or equivalently, relative to a reference point in the focal plane). The determination of this scale factor (the \emph{absolute} calibration) can be then achieved through the observation of standard sources with known brightness.

The partially overlapping \emph{ubercalibration} procedure was first developed and applied to the SDSS imaging data~\citep{Padmanabhan2008}. 
Further investigations of the method are reported in~\citep{holmes2012designing}, where it has been shown that quasi-random observing strategies provide more uniform coverage on the focal plane with respect to regular and semi-regular patterns.
Applications of the method by ground telescopes includes the PS1 survey~\citep{Schlafly2012}.
Preliminary studies applied to a space mission have been carried out in~\citep{Markovic2017}.

In this work, we first show possible ways to generalize the method outlined in~\citep{holmes2012designing}: we show how to reconstruct the response function using a generic two dimensional function basis, we introduce the possibility to further sectorize the focal plane to account for different macro-detectors, and we derive a rigorous procedure to accurately estimate the calibration uncertainty. We then quantify the performance for generic spectro-photometric surveys.
Synthetic simulations of in-flight self-calibration surveys are produced to study the accuracy on the inferred instrument response function (and source rates) under different scenarios.

The method described in this paper may be adopted by upcoming or future galaxy surveys to characterise their in-flight self-calibration, in order to consistently infer a parametric instrument response function using real calibration data, or to plan their in-flight calibration with simulated data.

This paper is organized as follows: section~\ref{sec:synt} describes the simulation of the synthetic calibration survey; section~\ref{sec:response} illustrates the general relative response functions used in this work; section~\ref{sec:inference} describes the minimization procedure to infer the parameters of the instrument response function and the source rates; section~\ref{sec:test} reports the results obtained in mock-up tests.

\section{The synthetic calibration survey}
\label{sec:synt}
Synthetic calibration surveys are simulated for studying the features of the in-flight self-calibration method.
The elements of the survey simulation are:
the sources entering the sky catalog, detailed in section~\ref{sec:sources}, the sky catalog, described in section~\ref{sec:sky}, the exposures, illustrated in section~\ref{sec:exposures}, the observations resulting from the exposures, described in section~\ref{sec:obs}.

\subsection{Sources}
\label{sec:sources}
The self-calibration procedure usually relies on bright stars, for both their high signal-to-noise ratio and their almost point-like detection over few pixels. In the case of slitless spectroscopy, bright stars are also employed because of the relatively negligible spectra cross-contamination from fainter neighboring sources and the well defined extraction aperture correction needed to derive the total flux from the extracted spectrum for each object.

In our synthetic sky catalogues, a calibration \emph{source} is described by three variables:
\begin{itemize}
\item the \emph{position} in the sky, identified by the standard Cartesian coordinates $(\xi, \eta)$ under a flat-sky approximation;
\item the source \emph{intrinsic count rate} $r$, i.e. the ideal number of detection counts per second in the instrument, due to the source.
\end{itemize}

\subsection{The sky catalog}
\label{sec:sky}

\begin{figure*}[tpb]
\centering
\includegraphics[width=.41\textwidth]{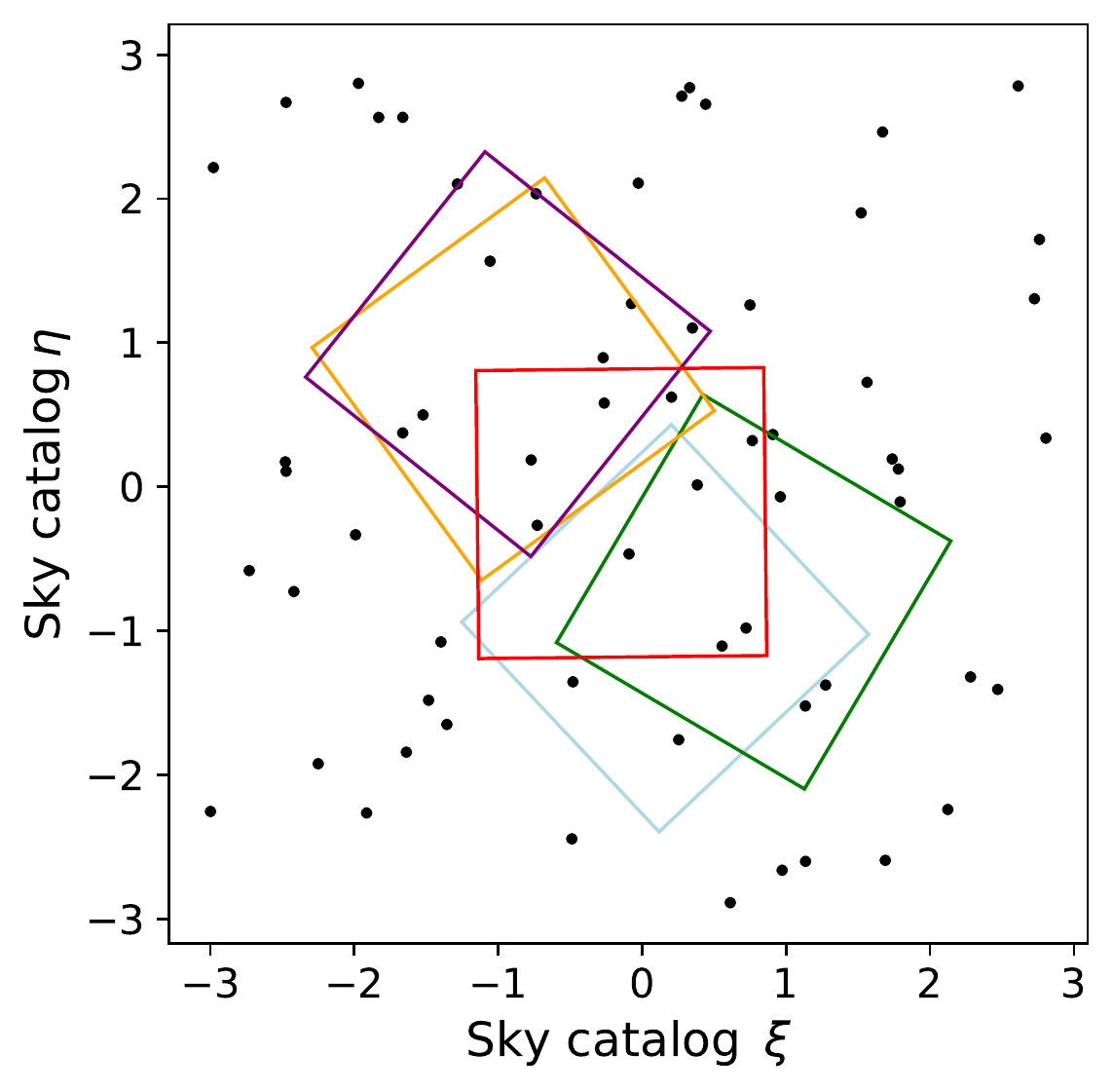}\hfill
\includegraphics[width=.44\textwidth]{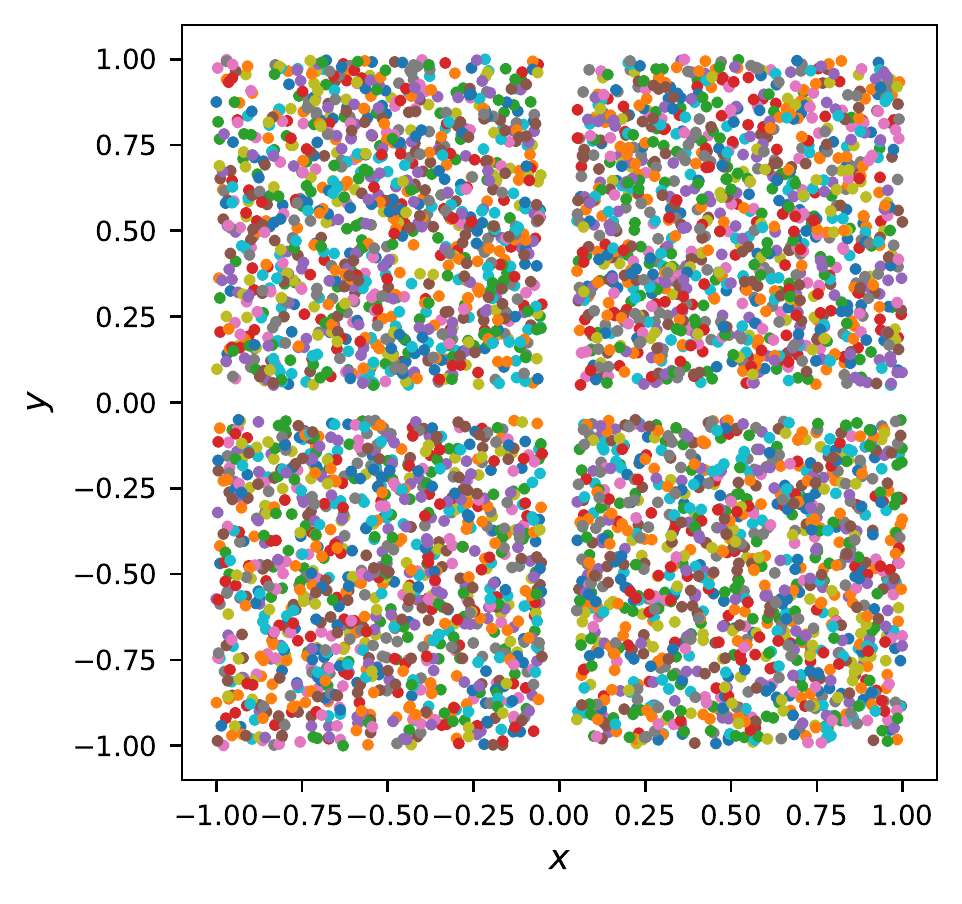}
\caption{\emph{Left}: Illustration of mock-up self calibration survey with an average of 7 sources per field of view and 5 exposures. 
\emph{Right}: Graphical visualization of the total observations in the focal plane, obtained with a sky catalogue with an average of 70 sources per field of view and a set of 60 exposures. The gap between the detector sectors is clearly visible.
}
\label{fig:observations}
\end{figure*}

\begin{figure*}[tpb]
\centering
\includegraphics[width=.43\textwidth]{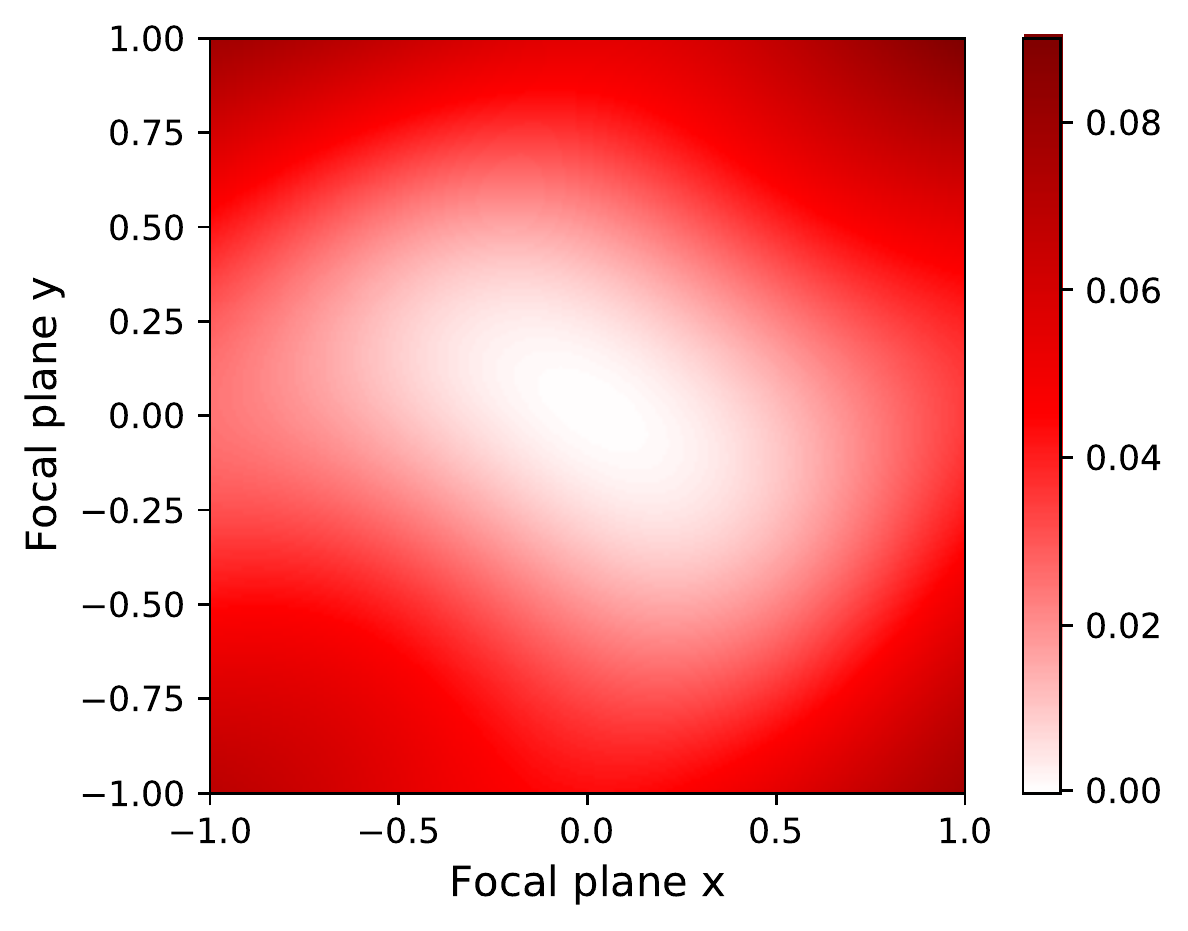}\hfill
\includegraphics[width=.43\textwidth]{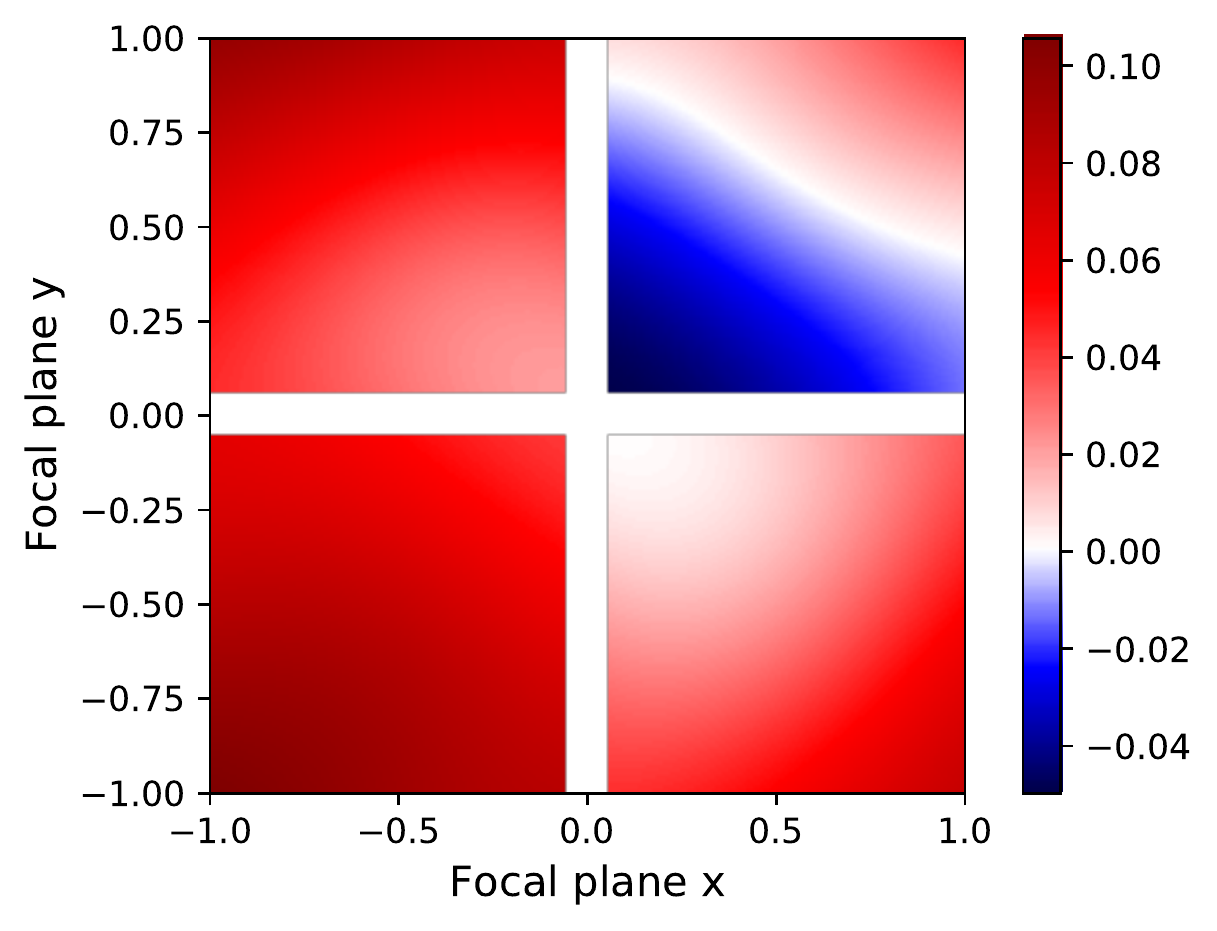}

\caption{\emph{Left}: Graphical visualization of the mock-up response function $\resp$ in the focal plane, in the case with one unsegmented detector; to enhance the contrast, the quantity $1-\resp$ is plotted. The values of $\resp$ in the four corners of the focal plane are approximately $0.922$ (top-left), $0.910$ (top-right), $0.924$ (bottom-right), $0.931$ (bottom-left).
\emph{Right}: Graphical visualization of $1-\resp$ in the case with four detector sectors, each with a different gain $\gs$ in each sector. The sector gains $g_s$ are respectively $(0.98, 1.05, 0.96, 1)$.
}
\label{fig:1-resp-focalplane}
\end{figure*}

The synthetic \emph{sky catalog} is a collection of sources, described by the sets $\{ \xi, \eta, r \}_k$ where $k$ is the source index.
The positions in the synthetic sky catalog are represented in the flat-sky approximation, which is appropriate for a few partially overlapping exposures over a total of few square degrees.

The origin of the sky coordinate and the orientation of the two Cartesian axes are arbitrary. The scale of the coordinate system is chosen such that a unit is half the size of the focal plane edge.

\subsection{Exposures}
\label{sec:exposures}

In our synthetic simulations, an \emph{exposure} is defined by four variables:
\begin{itemize}
\item a telescope \emph{pointing} in the sky, described by the pair of sky coordinates $(\xi, \eta)$;
\item an \emph{orientation} angle $\theta$;
\item the exposure \emph{time} $t$.
\end{itemize}
The set $\{ \xi, \eta, \theta, t \}_i$ describes the $i$-th exposure in the calibration synthetic survey.
The exposure is geometrically modelled as a square in the sky:
the pointing coincides with the geometric center of the square; the rotation of the square (with respect to the sky coordinate system) is $\theta_i$. 
In this work, the centers of the telescope pointing are extracted randomly in the central sector of about one third of the simulated sky area; the telescope orientation angles are also extracted randomly.
Details are provided in section~\ref{sec:mockup} and~\ref{sec:validation}.

The \emph{focal plane coordinate system} origin is defined to be the geometric center of the focal plane, with the Cartesian axes parallel to the edges of the focal plane.
The position in the focal plane coordinate system is then represented by a pair of coordinates, denoted by $(x,y)$.
The focal plane coordinates $(x,y)_i$ of a source with sky coordinates $(\xi, \eta)_k$,
observed in an exposure with pointing $(\xi, \eta)_i$ and orientation $\theta_i$, are derived with a roto-translation:
\numparts
\begin{eqnarray}
\label{eq:transform-x}
x_i &=& + (\xi_k - \xi_i) \cos \theta_i  \, +  \, (\eta_k - \eta_i) \sin \theta_i \\
\label{eq:transform-y}
y_i &=& -(\xi_k - \xi_i) \sin \theta_i  \, +  \, (\eta_k - \eta_i) \cos \theta_i
\end{eqnarray}
\endnumparts
In the case of a photometric survey, the weighted centers of luminosity of the source images can naturally be used as position coordinates. 
For a slitless spectroscopic survey, the  mid-point of the first order spectrum integrated over a proper wavelength range can be used as position coordinate ~\citep{Markovic2017}.

We define the domain of the focal plane coordinates to lie between $-1$ and $1$, i.e. $x \in [-1,1]$, $y \in [-1,1]$. The same scale is therefore used for the focal plane coordinates and the sky coordinates (see figure~\ref{fig:observations}).

\subsection{Observations}
\label{sec:obs}

The \emph{expected counts} $\meanik$ for the $k$-th source in the $i$-th observation is  obtained as the product of the intrinsic count rate of the source $\rate$, the exposure time $t_i$, and the assumed instrument response function $\resp$ evaluated in the focal plane coordinates $\fpxyk$,

\begin{equation}
\label{eq:meancnt}
\meanik  = f \fpxyk \, \rate  \, t_i \, .
\end{equation}
The \emph{response function} $\resp$ models the overall light collecting efficiency of the entire telescope and the spectro-photometric instrument, including variations in its optics.
In the synthetic survey simulation, $\resp$ is provided as input.
In this work, we employed three levels of complexity on top of the response function.
In the first level, we assume a single detector with a uniform gain over the whole focal plane; in the second level, the detector is segmented in four equal parts separated by small gaps; in the third complexity level, we account for small instrumental differences in the global performance of the four detectors. Each detector is parametrized with its own multiplicative scale factor called \emph{gain}. Examples of response functions used in the tests are given in figure~\ref{fig:1-resp-focalplane}.

The \emph{observed count} of the $k$-th source in the $i$-th observation is denoted by $\cnt$.
In general, the observed count receives contributions from both the signal and the noise.

The method presented in this work relies on the assumption that the noise is known with sufficient accuracy, either from a noise model or from data driven methods. A complete noise model includes variations of the noise across the focal plane and the detector elements, and over time.

In our simulations, we simply model the noise with a uniform term $n_i$ (for the $i$-th exposure) across the focal plane.
The contribution of $\noise$ can be parameterized as the sum of a constant plus a linear term in the exposure time.
This noise model is adequate for detectors whose parameters are known either from the manufacturers or from ground calibrations.
Implementing more realistic noise models is beyond the scope of this work.
Nevertheless, the inference procedure implemented in section~\ref{sec:inference} is suitable for more sophisticated noise models, as long as the noise mean and variance are known.

The observed $\cnt$ is sampled performing an extraction from a Poisson distribution with mean $ \meanik + \noise$ and subtracting the noise term from it, as follows:
\begin{equation}
\label{eq:extraction}
\cnt = {\rm Poisson}[ \meanik + \noise] - \noise \, .
\end{equation}
The model simulates the effect of a known stochastic noise
which is subsequently subtracted (e.g. in a post-exposure processing phase).
The net effect of $\noise$ in equation~(\ref{eq:extraction}) is an increase of the fluctuations of the random variable $\cnt$, which is ultimately distributed as a Poissonian distribution with expected value $\meanik$ and variance $\meanik + \noise$.

The inference procedure implemented in this work is based on a $\ki$ statistics; therefore, the method relies on the assumption that the count distribution is well described by a Gaussian.
In this work, we consider as usable sources for the in-flight calibrations those with counts in the range $10^4-10^6$, on top of which we include a noise $\noise$ of order $10^3$.
With these counts, the random variable distribution $\cnt$ in~(\ref{eq:extraction}) is well approximated by a Gaussian of mean $\meanik$ and sample variance 
\begin{equation}
\sig = \cnt + \noise \, ,
\end{equation}
thus justifying the use of a $\ki$-based method for the inference of the response function.
 
The self-calibration procedure provides a method for the statistical inference of the (a priori unknown) response function $\resp$. 
A key feature of the self-calibration method is that the inferred response function can only be determined up to a uniform scale factor.
This can be understood by noting the \emph{degeneracy} between the scale of the response function
(i.e. a multiplicative factor in $f$) and the intrinsic source rates $\rate$ in equation~(\ref{eq:meancnt}):
the detection is only sensitive to their product.
A scenario where all the sources are fainter by a common scale factor provides the same expected counts as a scenario where the instrument response function is uniformly lower by the same scale factor.

The degeneracy can be handled by interpreting the response function as relative to an arbitrary reference point in the focal plane.
The function $\resp$ thus models the ratio of the instrument response to the response in the reference point ${(x,y)}_{\rm ref}$:
 \begin{equation}
 f^{\rm{relative}} (x,y) := \frac {f^{\rm{absolute}} (x,y)} {f^{\rm{absolute}} {(x,y)}_{\rm{ref}}} \, . 
 \end{equation}
The relative response function clearly returns one in the reference point.
We choose the reference point as the center of the focal plane, i.e. in the coordinate pair $(x=0, y=0)$.

Throughout the paper, we drop the `relative' and `absolute' specifications in the notation of $\resp$:
$\resp$ always refers to the \emph{relative} response function, unless otherwise specified.
 
The \emph{synthetic calibration survey} is then described by the collection of the sets $\{\cnt, \sig, \fpxyk, t_i\}$, spanning over $k$ sources and $i$ exposures.
 
\section{The parametric relative response function}
\label{sec:response}

The reconstructed relative response function, denoted by $\rreco$, is parametrized in order to account for a smooth variation due to the telescope optics $\reco$, on top of possible discontinuous effects due to the use of detectors with slightly difference performances in the different sectors. 

The function $\rreco$ is defined in the domain $x \in [-1,1]$ and $y \in [-1,1]$ and is represented as:
\begin{equation}
\label{eq:expansion2}
\rreco =  \sum_{\ell=0}^N \ql \, \basis (x,y) \,  \sum_{s} g_s \Thetasxy \, .
\end{equation}
The first sum is a linear combination over a basis in a space of two dimensional continuous functions:
\begin{equation}
\label{eq:recocontinuous}
\reco =  \sum_{\ell=0}^N \ql \, \basis (x,y) .
\end{equation}
The function $\basis$ is the $\ell$-th element of the basis set $\{ \basis (x,y) \}$ and the \emph{coefficient} $\ql$ is the corresponding element of the coefficient vector $\qvec$.

The choice of the basis is  arbitrary: the method for inferring the coefficients $\qvec$ illustrated in this work is general with respect to the choice of the basis.
Nevertheless, some of the special functions of mathematical physics are particularly suited as basis set.
In this work, we tested three sets of basis:
the set of powers, the Legendre polynomials, and the Fourier basis.
The construction of these bases is detailed in~\ref{sec:expansion}.

The second sum of~(\ref{eq:expansion2}) is extended to all the  sectors, and $\Thetasxy$ is the projector on the $s$-th sector: $\Thetasxy$ equals one if the coordinate $(x,y)$ belongs to the detector sector $s$, and it is zero elsewhere. 
The response in each sector is further multiplied by a scale factor $\gs$, that we generically name \emph{sector gain}.
The set of the gains $\{\gs\}$'s is conveniently represented as a vector $\gvec$. 
The purpose of the $\{\gs\}$'s is to parametrize the slightly different performances of each detector which may be due to its intrinsic efficiency and signal amplification. 

Following the conventions outlined in section~\ref{sec:obs},
the relative response function must return unity in the origin of the focal plane coordinates.
Once the basis set $\{ \basis \}$ is chosen, the normalization of the relative response function $\hat f  (0,0 | \qvec, \gvec) = 1$
can be converted into a constraint on the coefficients $\{ \ql \}$'s and on the gain of the central sector $g_c$:
\begin{equation}
\label{eq:normalizationql}
g_c\sum_{\ell=0}^N \ql \basis (0,0) = 1 \, .
\end{equation}
The normalization constraint must always be satisfied,
regardless of the basis.
Without loss of generality, we set the gain in the central sector $g_c$ to one: all the gains in the other sectors are relative to the gain in the central sector. In case there is only one single sector, equation~\ref{eq:expansion2} reduces to equation~\ref{eq:recocontinuous}.

\section{Statistical inference}
\label{sec:inference}

In the in-flight self calibration method, the inference of the set of source rates $\{ \rate \}$'s, the relative response coefficients $\{ \ql \}$'s, and the relative gains $\{ gs \}$'s is obtained by comparing each observed count $\cnt$ against its expected count.

As described in section~\ref{sec:synt}, we are employing a synthetic calibration survey, and we model an observation by the set $\{\cnt, \sig, \fpxy, t_i\}$;
nevertheless, the method described here can be employed with no modifications in a realistic survey, e.g. using the detected counts for $\cnt$ and a combination of data driven methods and simulations for $\sig$.

One of the advantages of the method is that it can be used  without any additional prior information about the $\{ \rate \}$'s, the $\{ \ql \}$'s, and the $\{ \gs \}$'s:
both in the synthetic calibration survey output and in a real survey, the intrinsic source count rates and the response function are initially unknown.

A test statistics must be chosen in order to perform a statistical parametric inference.
Since the in-flight self calibration method deals with counts, a choice could have been made towards a likelihood inference, based on Poisson statistics.
However, given the high-statistics of the observed counts, the Gaussian approximation is more than adequate,
and a $\kisq$ can be used instead of the likelihood.

Using a Neyman's $\kisq$ as test statistics, we can derive linear expressions for the best values of the $\{ \rate \}$'s, the $\{ \ql \}$'s, and the $\{ \gs \}$'s as detailed in~\ref{sec:iterative}.
Also, the value of the $\kisq$ at the minimum is an indicator of the goodness of the model.

The matrix of the second derivatives of the $\kisq$ with respect to the $\{ \rate \}$'s, the $\{ \ql \}$'s, and the $\{ \gs \}$ can be written in an explicit form.
The inverse of the second derivatives matrix, the covariance matrix, is used to estimate the statistical uncertainty on the rates $\{ \delta \rate \}$'s and on the reconstructed relative response function, $\delta \rreco$. The computation of the uncertainties is detailed in~\ref{sec:uncertainties}.

\subsection{The $\kisq$ of the in-flight self calibration method}
\label{sec:kisqminim}
 
A Neyman's $\kisq$ is used as the test-statistics and for the parameter inference of the source rates $\{ \rate \}$'s, the coefficients $\{ \ql \}$'s, and the relative gains $\{ \gs \}$'s of the parametric relative response $\rreco$. The experimental $\kisqe$ is
\begin{equation}
\label{eq:kisq}
\kisqe \, = \, \sumk \, \sumik{ \frac {\left [ \cnt  \, - \,  \rrecofpk  \rate t_i \right ]^2} {\sig} } \, .
\end{equation}
The sum iterates over each source ($k$ label) and over the $\nk$ exposures ($i$ label) where the $k$-th source has been observed.

In the numerator of equation~(\ref{eq:kisq}) the observed count $\cnt$ of the $k$-th source in the $i$-th exposure is compared to its expected value;
the expected (theoretical) count is the product $\rrecofpk \rate t_i$,
where the reconstruction function is evaluated at the focal plane coordinates of the observation.

The denominator of equation~(\ref{eq:kisq}) is the variance $\sig$ of the $i$-th observation of the $k$-th source.
In a real survey, $\sig$ can be estimated from data driven methods or Monte Carlo simulations.

As described in section~\ref{sec:obs}, the Gaussian approximation is valid and the use of a $\kisq$ is adequate.
At the $\kisq$ minimum, the experimental $\kisqe$~(\ref{eq:kisq}) follows a $\kisq$ distribution
with the number of degrees of freedom $\ndf$ given by the number of source observations,
minus the numbers of sources, minus the number of coefficients $\{ \ql \}$ except one, minus the number of detector sectors except one:
\begin{equation}
\label{eq:ndf}
\ndf \, \equiv \,  \sumk \, (\nk - 1) - \sum_{\ell=1}^N 1 \, - \sum_{s=1}1 \, .
\end{equation}

The values of the source intrinsic rates $\{ \rate \}$'s, of the relative response coefficients $\{ \ql \}$'s, and of the relative gains $\{ \gs \}$'s can be inferred by minimizing the $\kisqe$ in~(\ref{eq:kisq}).
The minimization must be subject to the additional constraint that the relative response $\rreco$ equals one in the focal plane origin.
In summary, the minimization of the $\kisqe$ is subject to the following conditions:
\begin{equation}
\label{eq:kisqminim}
\left \{ 
\begin{array}{l}
\dkisqdr = 0  \\
\\
\dkisqdq = 0 \quad (\ell = 1, \dots, N) \\
\\
 \sum_{\ell=0}^N \ql \basis (0,0) = 1 \\
 \\
\dkisqdg = 0 \quad (s \mbox{ not central})
\end{array}
\right.
\end{equation}

The minimum of the $\kisqe$ is given by the solution of the system~(\ref{eq:kisqminim}).
We choose to solve the system with an expectation-maximization  iterative procedure, whose implementation is detailed in~\ref{sec:iterative}.

The numeric solution of~(\ref{eq:kisqminim}) starts by initializing $\rreco$ to the uniform response: the coefficients $\{ \ql \}$'s are all set to zero, except
$q_0 = 1/w_0(0,0)$, and all the gains $\{ \ql \}$'s are set to one.
The iterative procedure then repeats the following four steps:
\begin{enumerate} 
\item The intrinsic source rates $\{ \rate \}$'s are estimated with equation~(\ref{eq:dkisqdrarrow}),
where the parameters of $\rreco$ are fixed at the previous iteration values.
\item The response coefficients $\{ \ql \}$'s ($\ell = 1, \cdots, N$) are updated with equation~(\ref{eq:dkisqdqarrow}), where the $\{ \rate \}$'s and the $\{ \gs \}$'s are fixed at the previous step values.
\item The response coefficient $q_0$ is updated following the normalization constraint (equation~\ref{eq:constraintq0}), where the other $\{ \ql \}$'s are fixed at the previous step values.
\item The gains $\{ \gs \}$'s are updated with equation~(\ref{eq:dkisqdgarrow}), where the $\{ \rate \}$'s and the $\{ \ql \}$'s are fixed at the previous step values. The gain of the central sector is always fixed to one.
\end{enumerate}

The iteration is stopped when the $\kisqe$, computed with~(\ref{eq:kisq}) after the fourth step,
differs from the $\kisqe$ computed in the previous iteration by less than a configurable amount (set by default to $10^{-3}$).

The \emph{statistical uncertainties} of the intrinsic source rates $\{ \drate \}$, of the response coefficients  $\{ \dql \}$, and of the gains $\{ \gs \}$ are estimated from the diagonal elements of the covariance matrix, computed as the inverse of the (halved) second derivatives matrix of the $\kisqe$. The computation is detailed in~\ref{sec:uncertainties}.

Specific tests, detailed in~\ref{sec:validation}, have been performed to validate the inference procedure.
The validation tests confirmed that the inference of the intrinsic source rates $\{ \rate\}$'s, the relative response coefficients $\{ \ql \}$'s, the relative gains $\{ \gs \}$'s, and the parametric reconstructed relative response function $\rreco$ provide unbiased estimates.
The test confirmed also that the minimum value of the experimental $\kisqe$ obtained by the iterative minimization procedure follows indeed a $\kisq$ distribution with $\ndf$ given by~(\ref{eq:ndf}).

In all our tests, the solution of the system~(\ref{eq:kisqminim}) converged with fewer iterations when using the Legendre polynomial basis, compared to the other bases.

\section{Test on mock-up response function with realistic conditions}
\label{sec:test}

\newcommand{\mad}{\mbox{MAD}}
\newcommand{\cad}{\mbox{CAD}}
\newcommand{\ufoseven}{\mbox{UF}(0.7\%)}
\newcommand{\ufofive}{\mbox{UF}(0.5\%)}
\newcommand{\ufoone}{\mbox{UF}(0.1\%)}

\begin{figure*}[tpb]
\centering
\includegraphics[width=.48\textwidth]{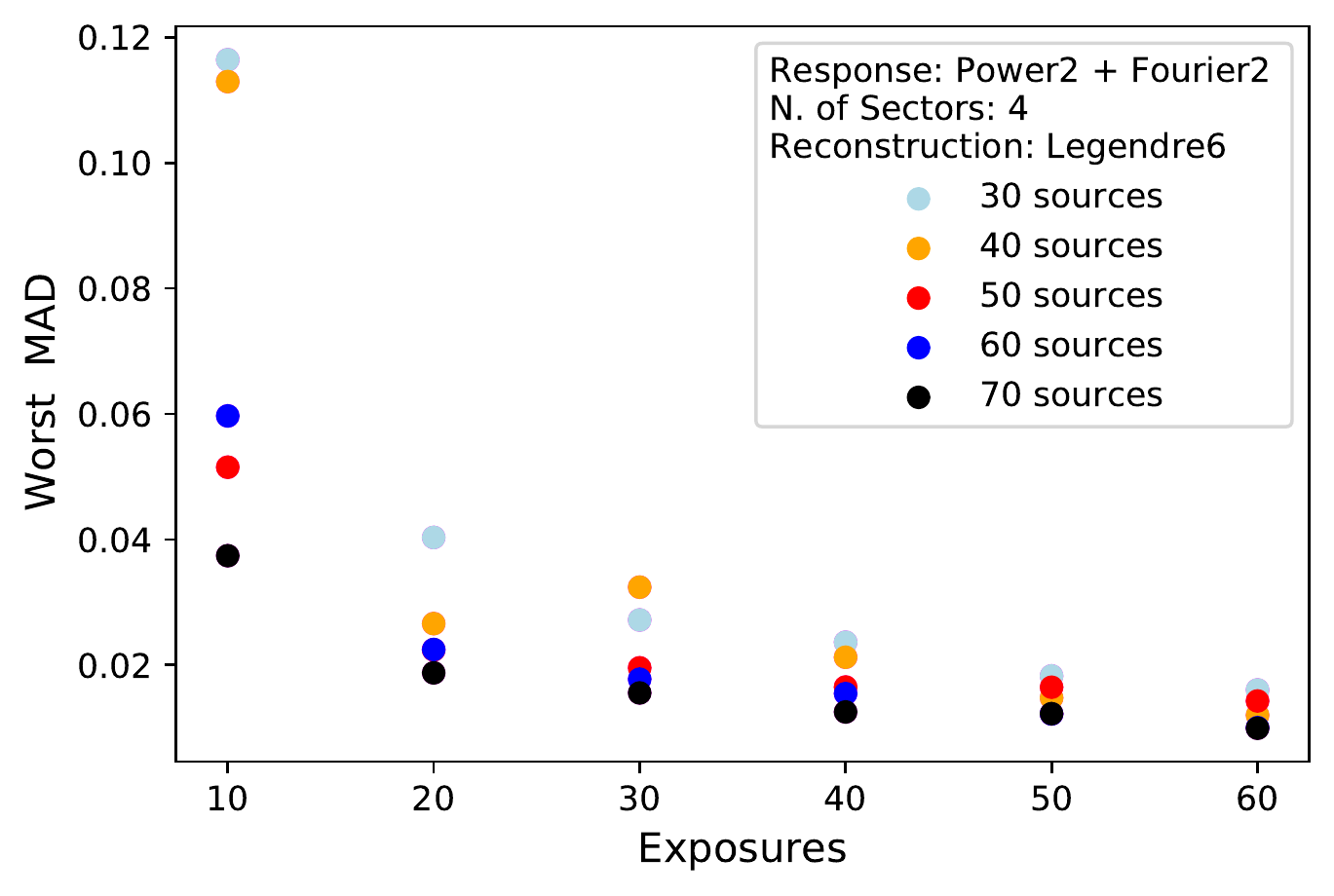}
\includegraphics[width=.48\textwidth]{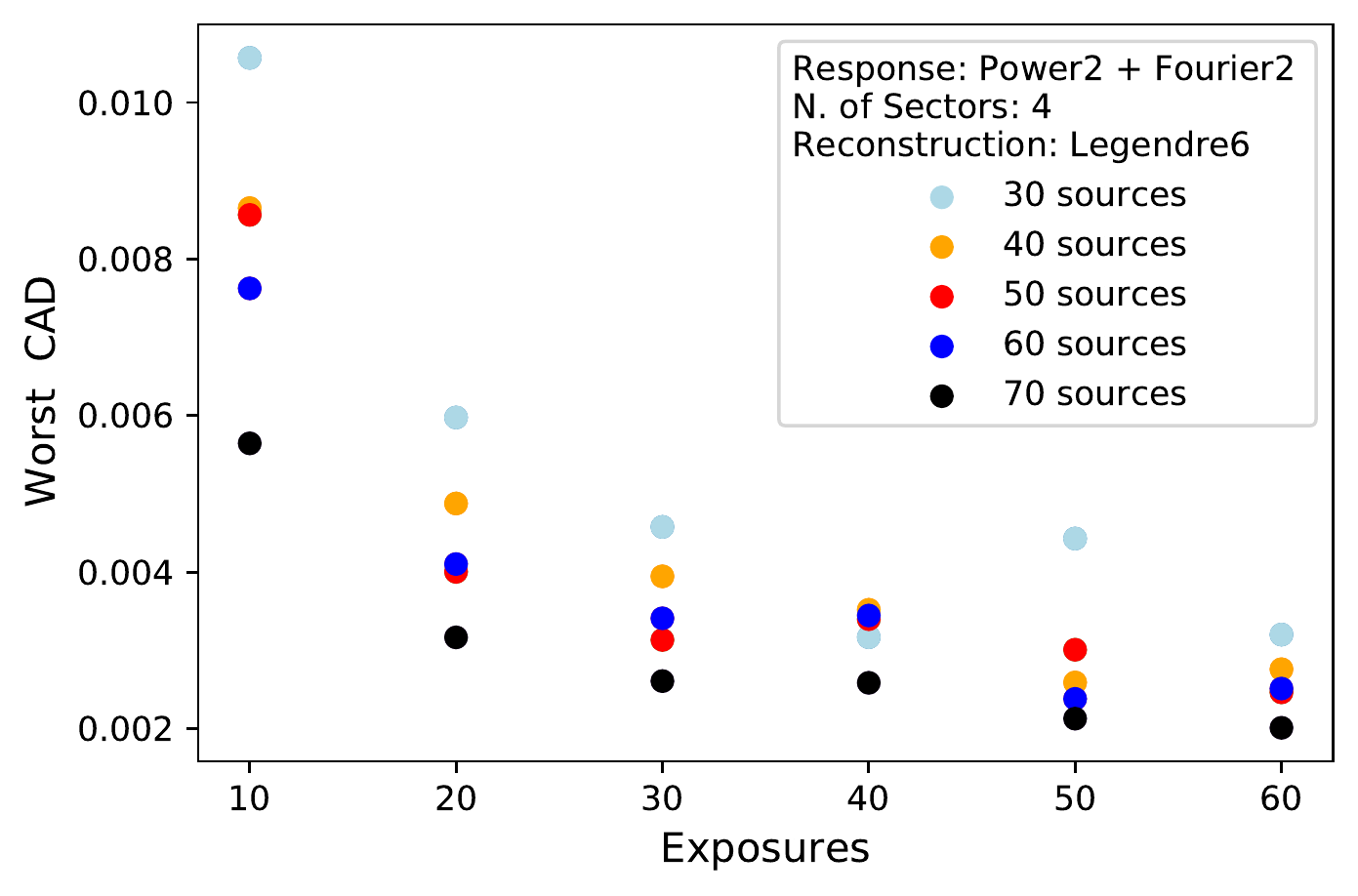}
\caption{Mock-up tests results: scatter plot of \mad\ (maximum absolute difference) and \cad\ (cumulative absolute difference), obtained in the realizations with the worst values, against the number of average sources in the field of view and the number of exposures. The reconstruction is performed with a Legendre polynomial basis, with maximum degree 6.
}
\label{fig:worst-MAD-CAD-leg6}
\end{figure*}

\begin{figure*}[tpb]
\centering
\includegraphics[width=.48\textwidth]{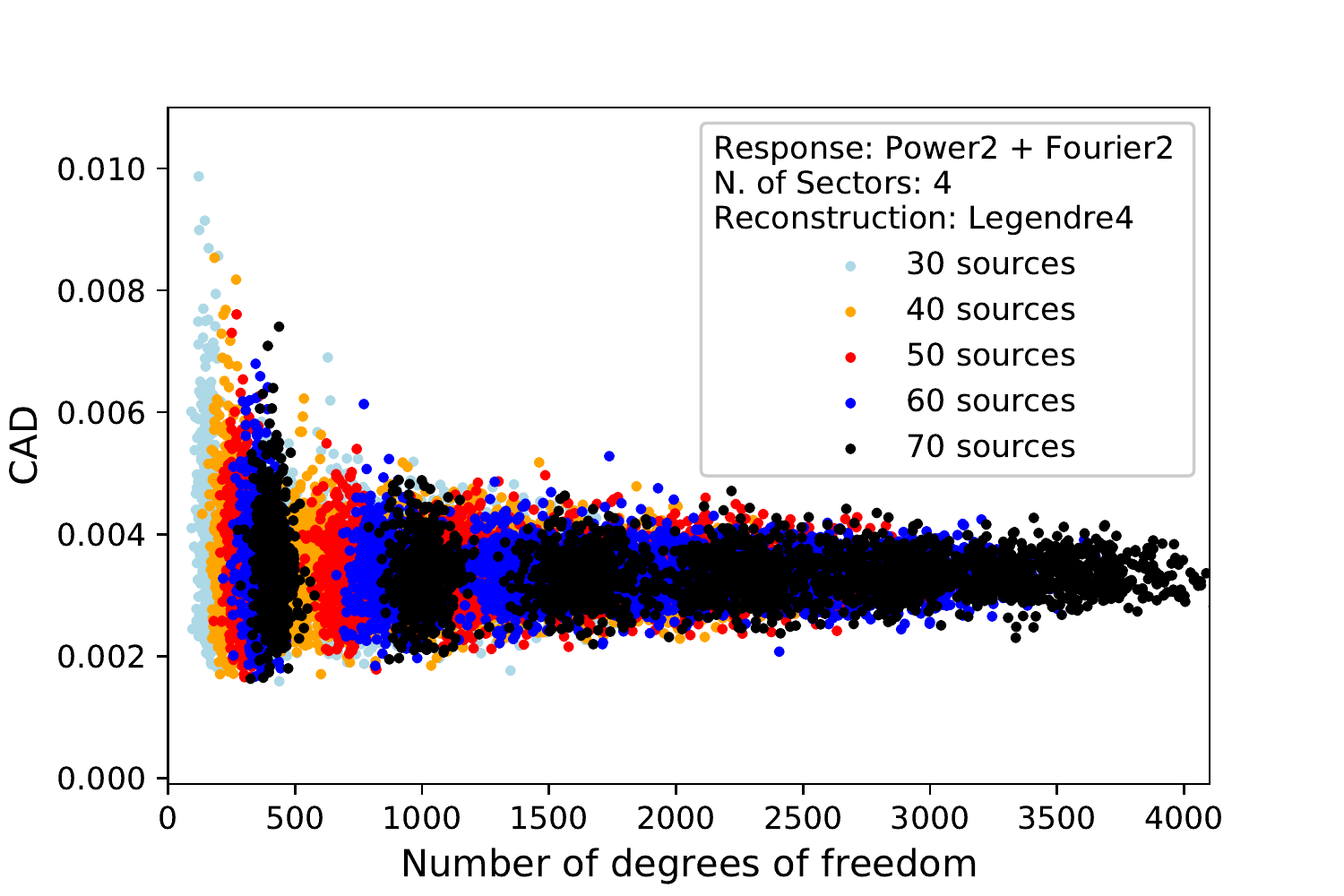}
\includegraphics[width=.48\textwidth]{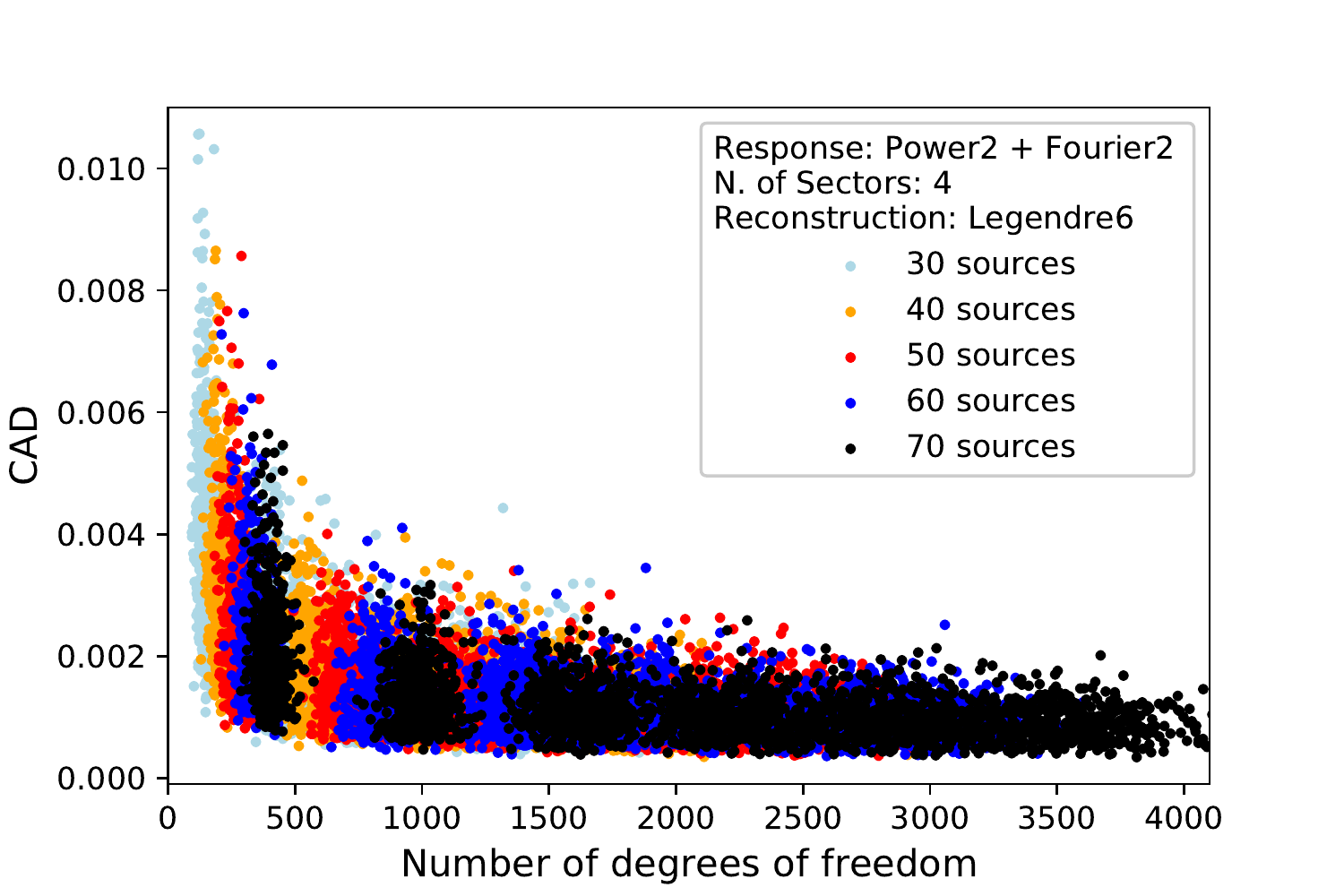}
\caption{Mock-up tests results: scatter plot of $\cad$ (cumulative absolute difference) against the number of degrees of freedom of the realization. The reconstruction is performed with a Legendre polynomial basis, with maximum degree 4 (\emph{left}) and 6 (\emph{right}).
}
\label{fig:tot-CAD-4567}
\end{figure*}

\begin{figure*}[tpb]
\centering
\includegraphics[width=.48\textwidth]{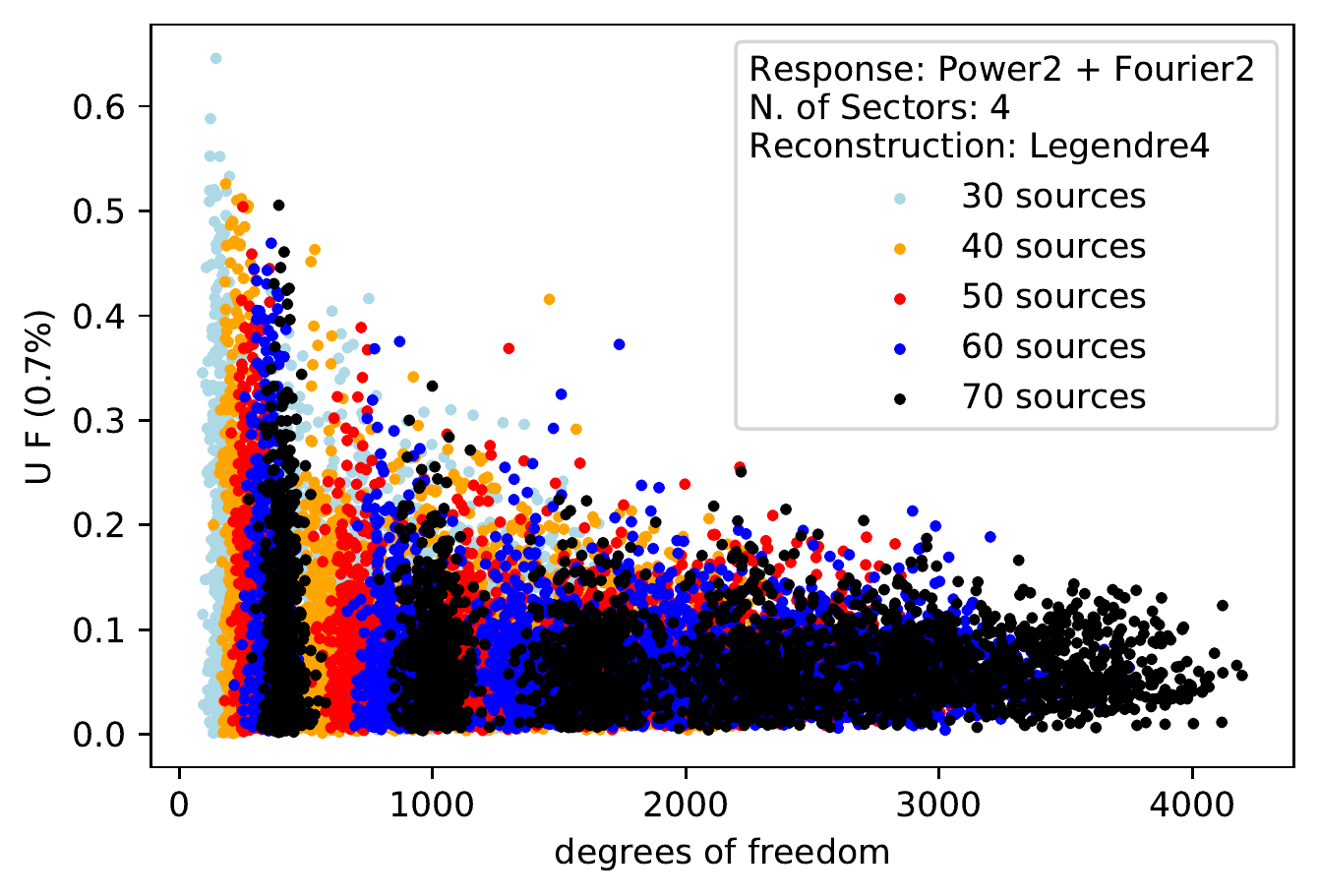}
\includegraphics[width=.48\textwidth]{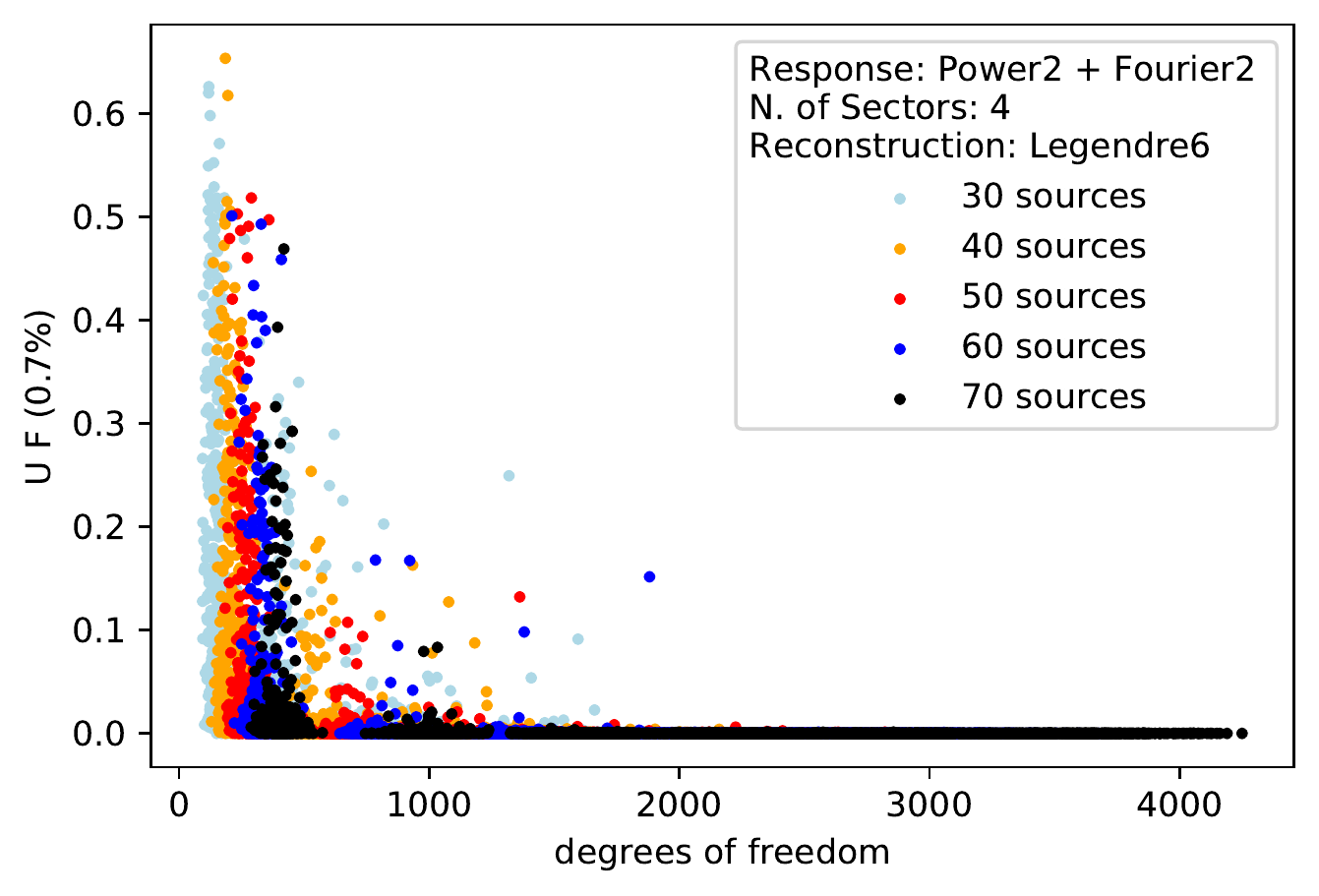}
\caption{Mock-up tests results: scatter plot of $\ufoseven$ (unusable fraction given a threshold of $0.7\%$) against the number of degrees of freedom of the realization. The reconstruction is performed with a Legendre polynomial basis, with maximum degree 4 (\emph{left}) and 6 (\emph{right}).}
\label{fig:tot-UF07-4567}
\end{figure*}

\begin{figure*}[tpb]
\centering
\includegraphics[width=.49\textwidth]{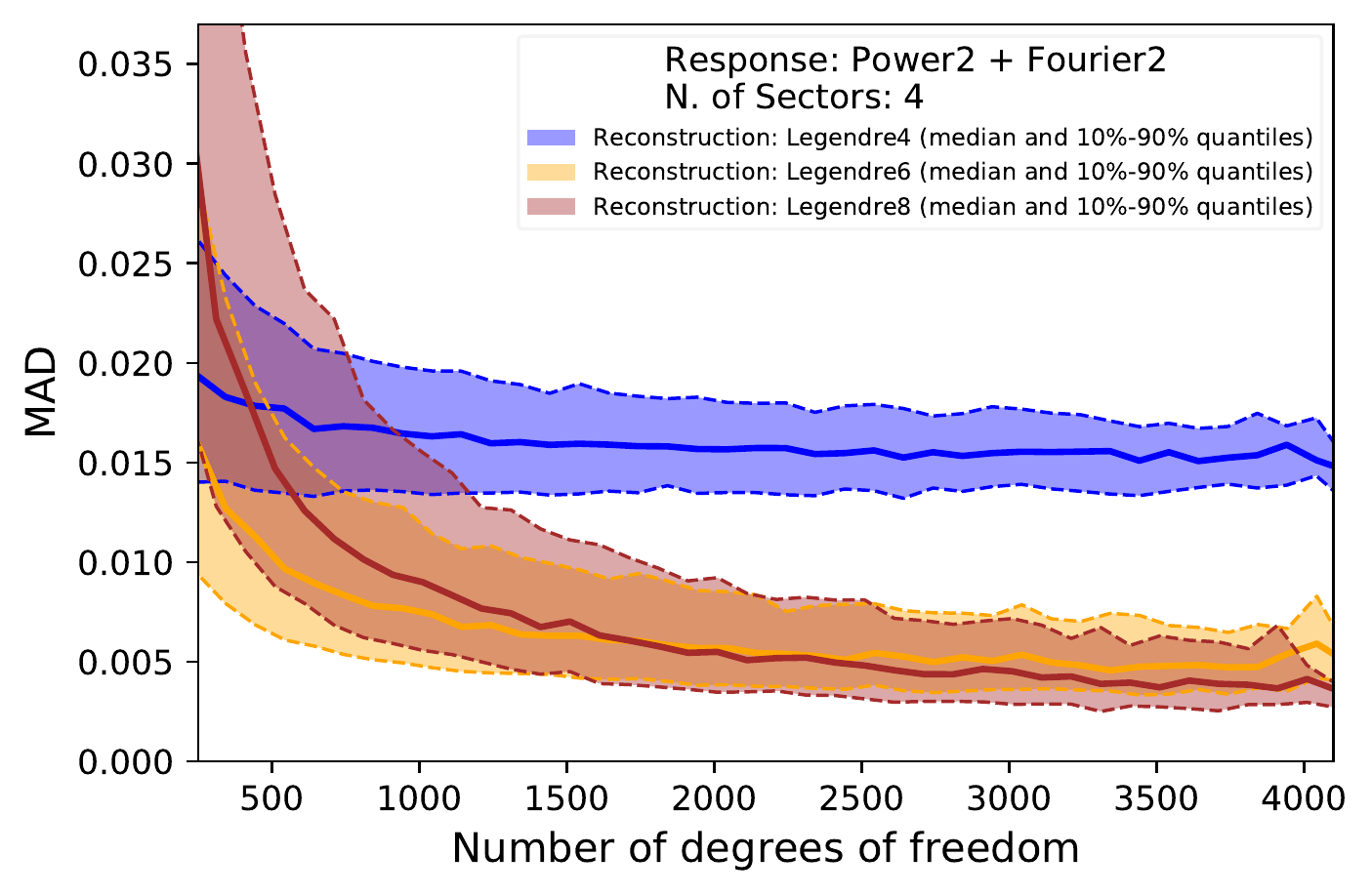}\hfill
\includegraphics[width=.49\textwidth]{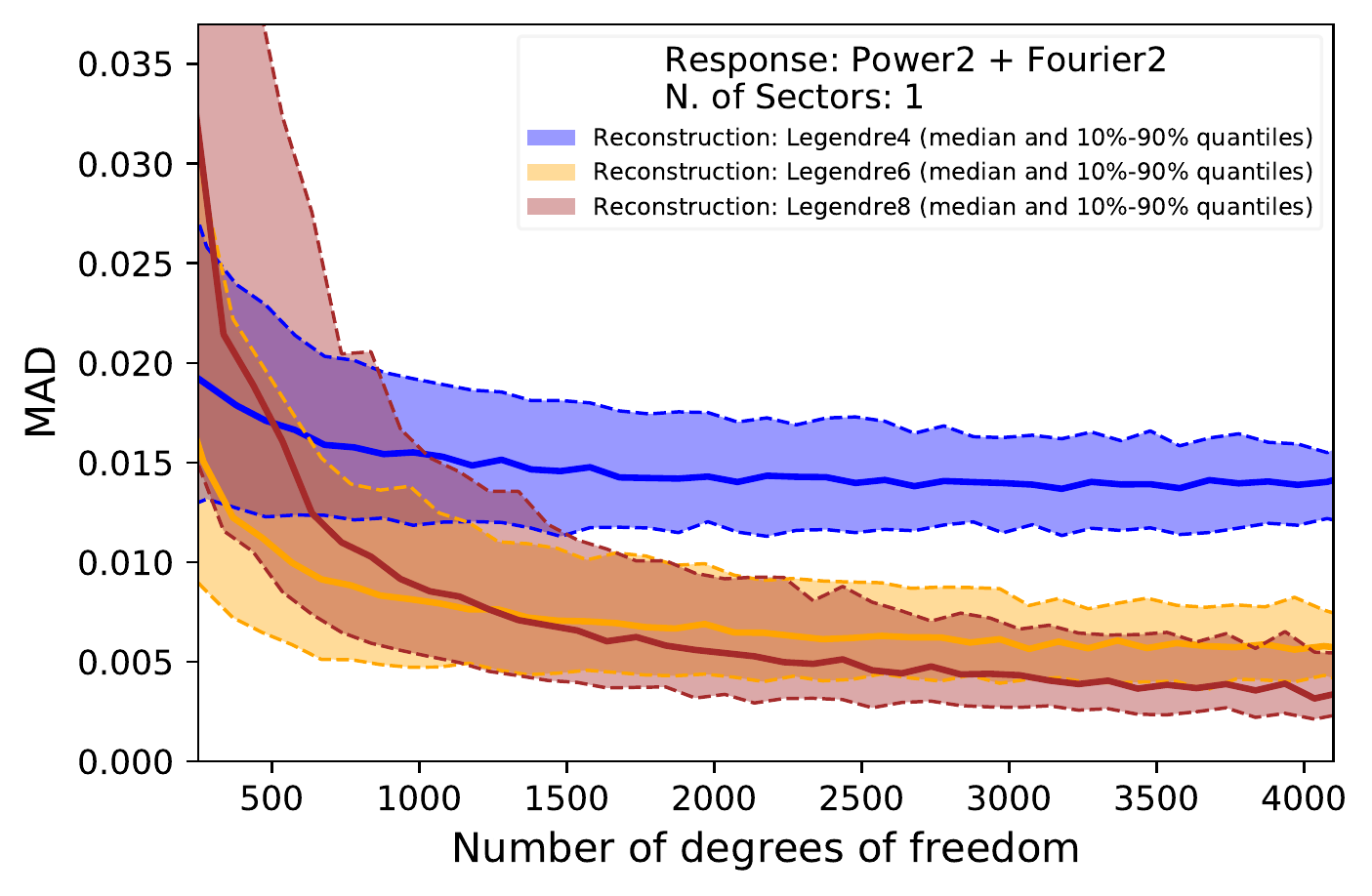}
\medskip
\includegraphics[width=.49\textwidth]{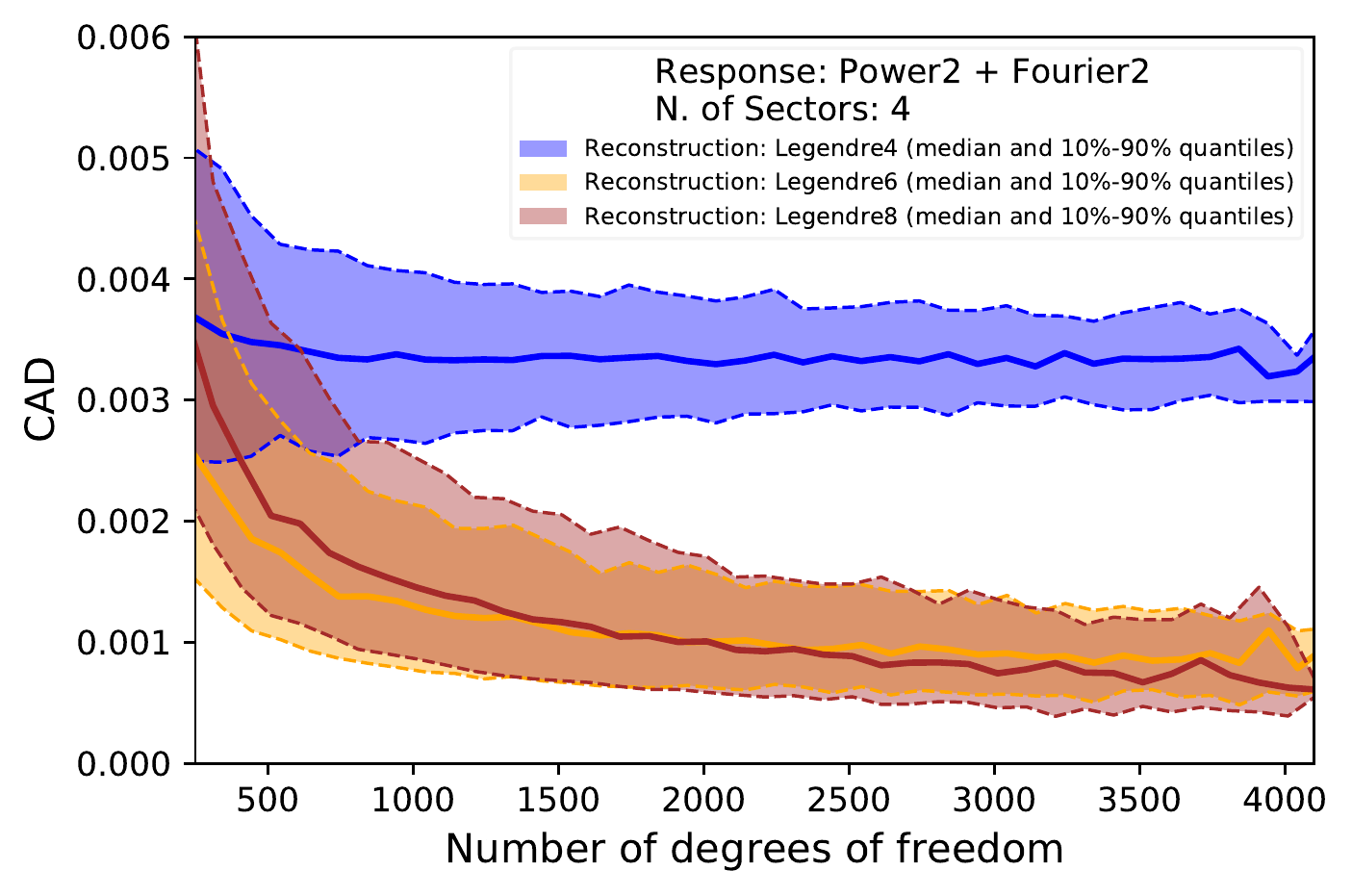}
\includegraphics[width=.49\textwidth]{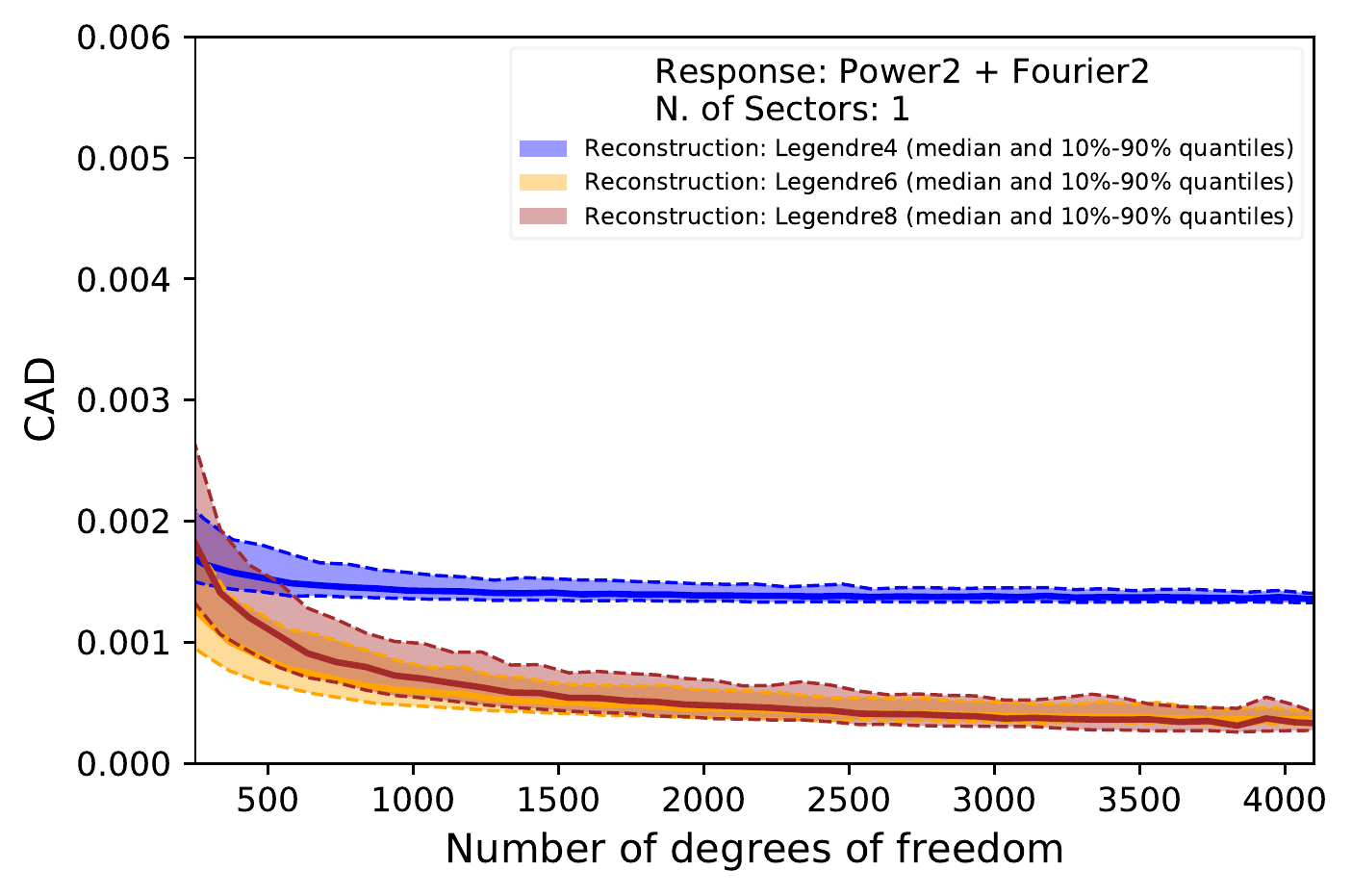}
\medskip
\includegraphics[width=.49\textwidth]{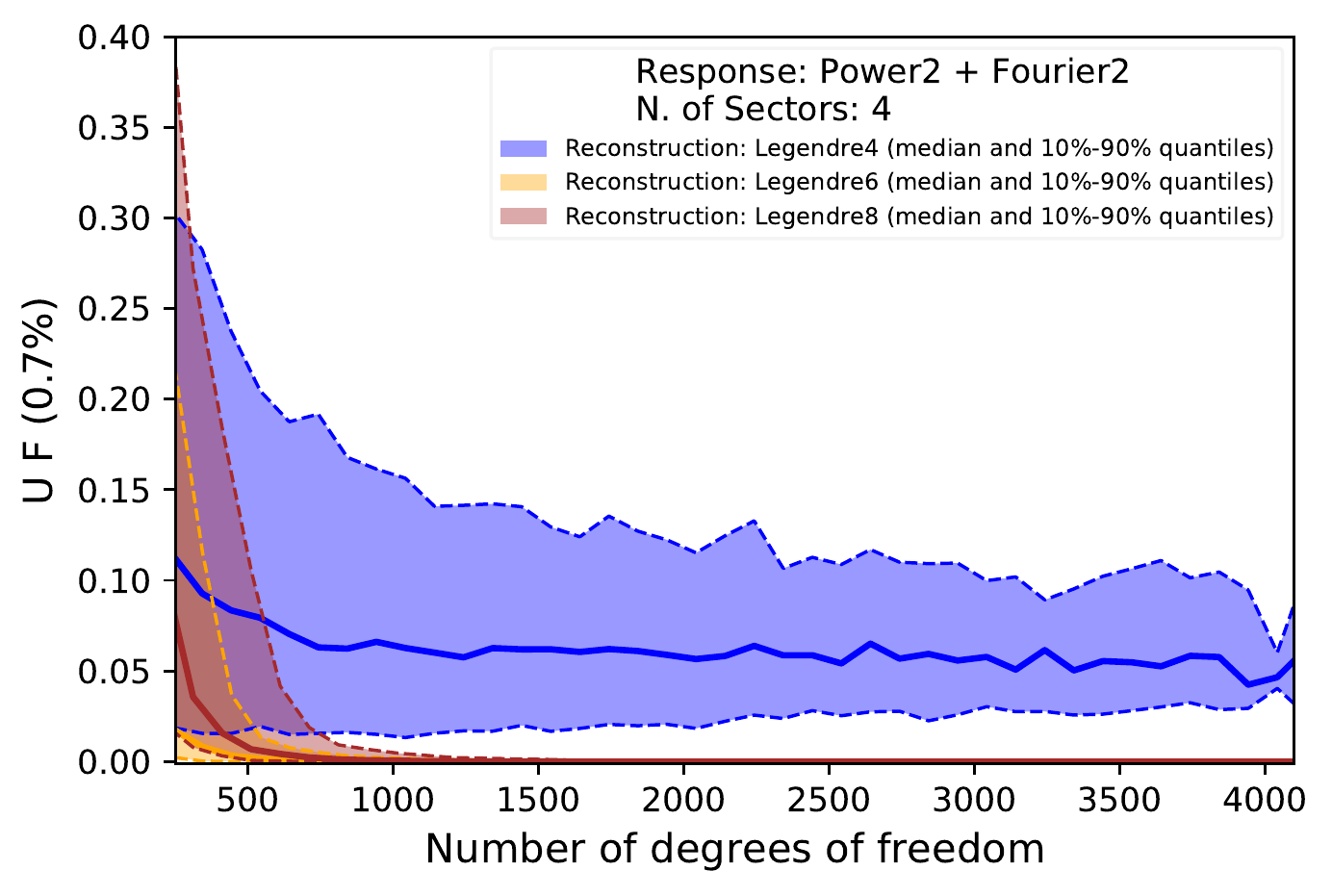}\hfill
\includegraphics[width=.49\textwidth]{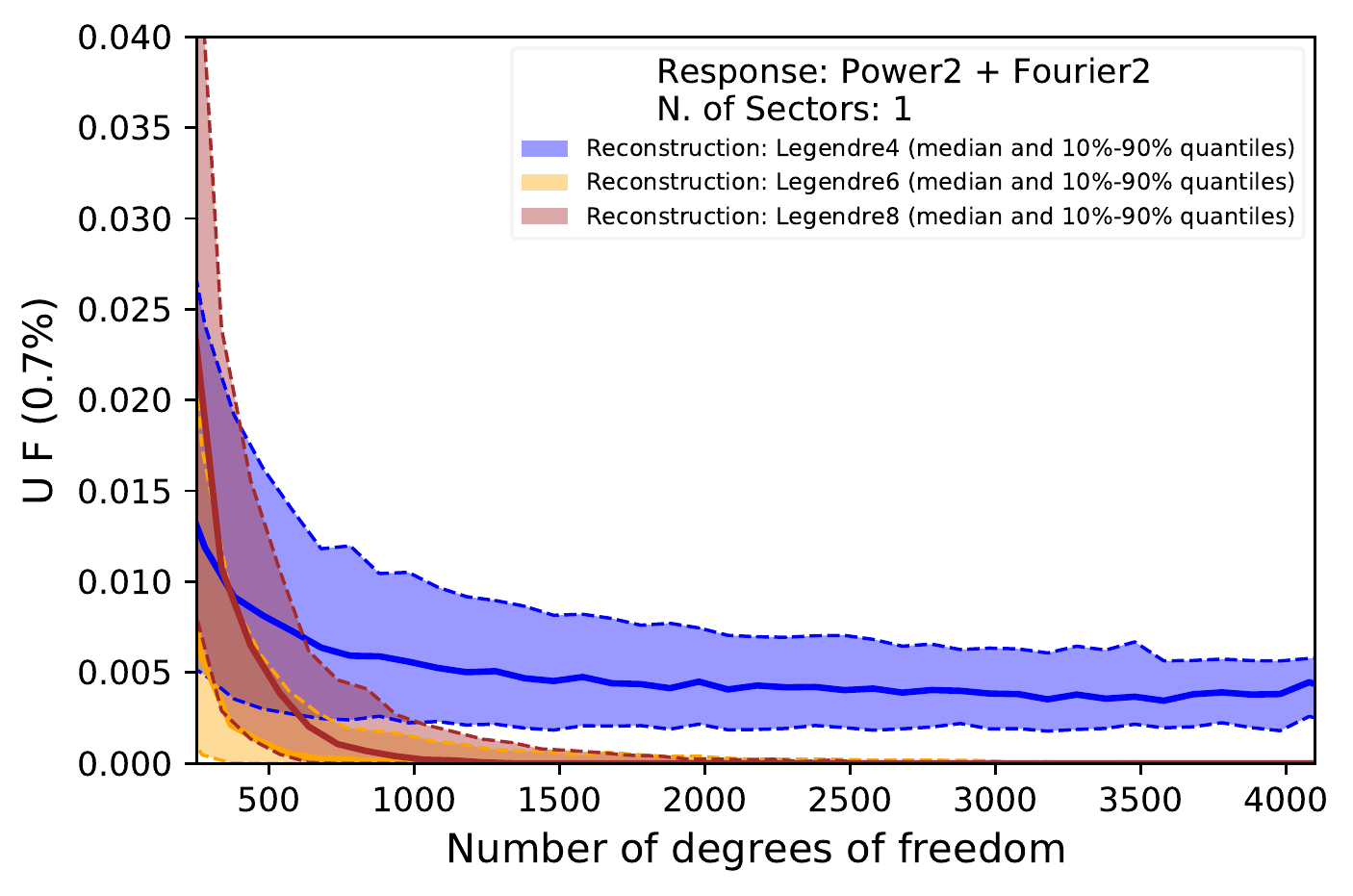}
\caption{Mock-up tests results: trends plots of the goodness metrics against the number of degrees of freedom of the realization and the Legendre basis maximum degree. The \emph{left} column refers to a detector with 4 sectors, each one with a different gain $\gs$ and with small gaps between sectors; the \emph{right} column refers to a single unsegmented detector over the focal plane.
Each median of the distributions is represented by the thick line. The distribution enclosed within the $10\%$ and $90\%$ quantiles is represented by the shaded area. 
The metrics are: $\mad$ (\emph{top}), $\cad$ (\emph{centre}), and $\ufoseven$ (\emph{bottom}). Note that the scale of $\ufoseven$ differs of a factor 10 in the two cases.
}
\label{fig:trend}
\end{figure*}

\begin{figure*}[tpb]
\centering
\includegraphics[width=.49\textwidth]{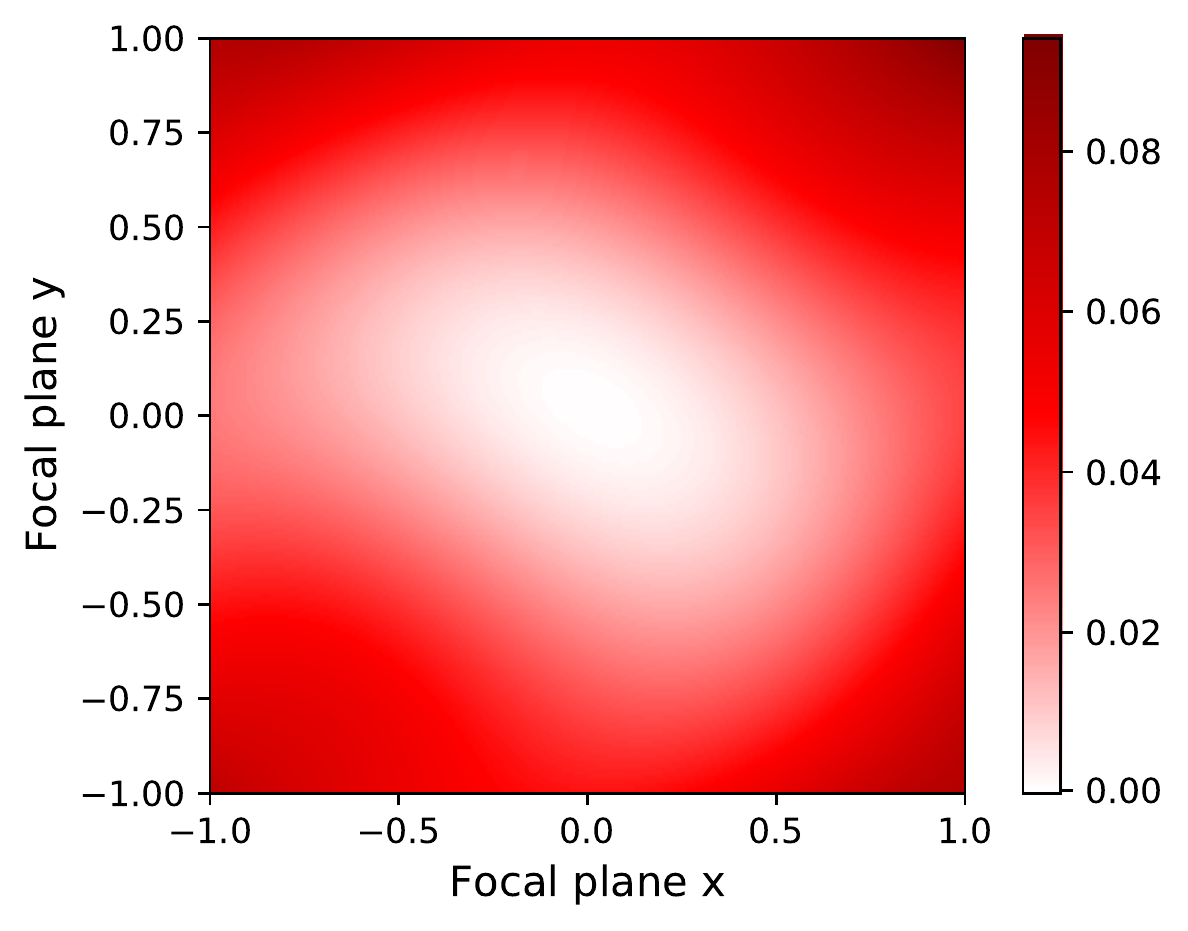}
\hfill
\includegraphics[width=.47\textwidth]{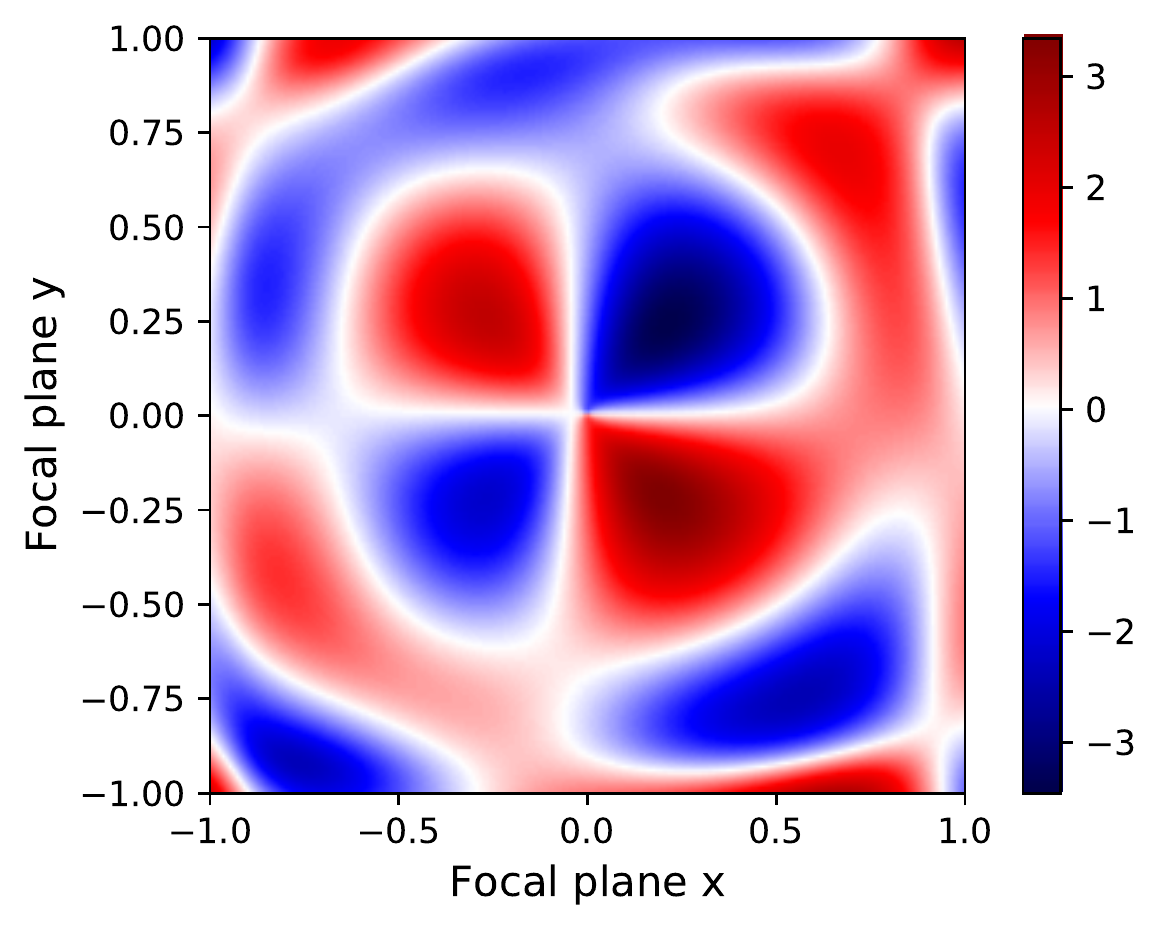}
\caption{\emph{left}: reconstructed $1-\reco$ in a mock-up test with 1 sector, using a Legendre reconstruction basis of maximum degree 6 ($\kisqe/ \ndf = 3927.59/3766 = 1.043$);
\emph{right}: corresponding residuals $(\resp - \reco )/ \delta \reco$ in the focal plane.
}
\label{fig:mock-residuals-1sector}
\end{figure*}

\begin{figure*}[tpb]
\centering
\includegraphics[width=.49\textwidth]{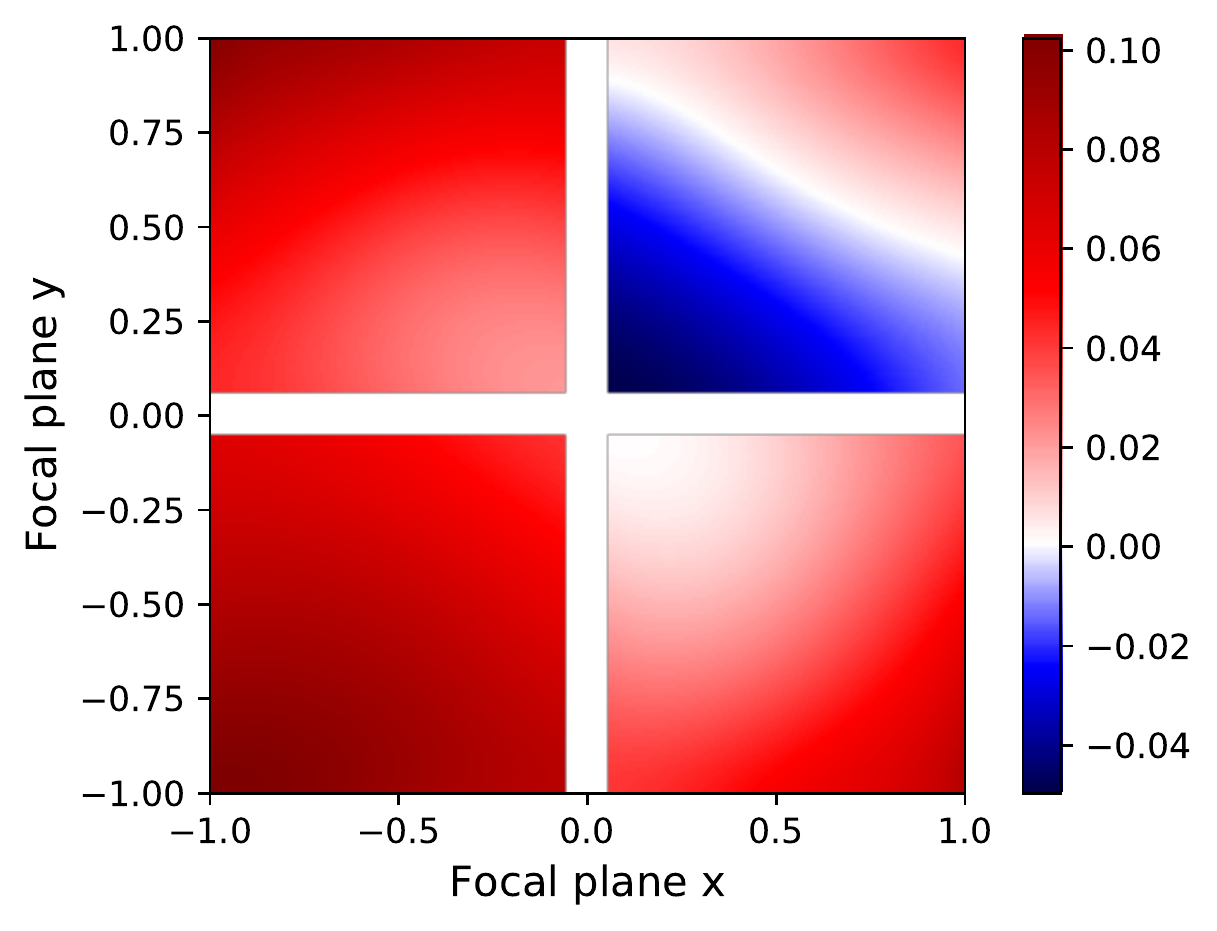}
\hfill
\includegraphics[width=.47\textwidth]{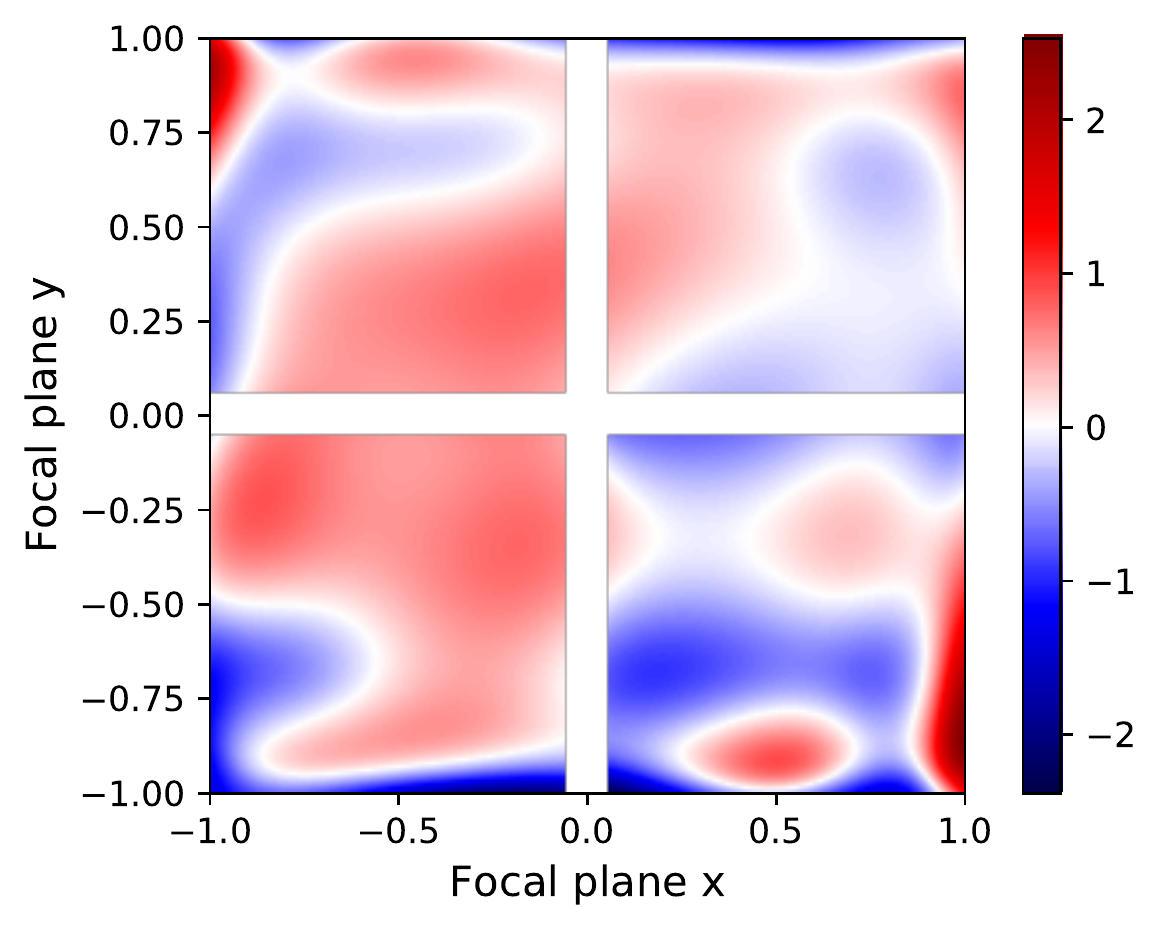}
\caption{\emph{left}: reconstructed $1-\rreco$ in a mock-up test with 4 sectors, using a Legendre reconstruction basis of maximum degree 8 ($\kisqe/ \ndf = 3228.66/3430 = 0.941$);
\emph{right}: corresponding residuals $(\resp - \rreco )/ \delta \rreco$ in the focal plane.
}
\label{fig:mock-residuals}
\end{figure*}

The framework presented in the previous sections can be adopted by upcoming galaxy surveys to characterize the in-flight self-calibrations, simulating distributions of sources, exposures, and response functions representative of realistic conditions.
The simulations can be used to evaluate the goodness of the instrument reconstruction function and choose among several self-calibration plans.
 
In this section, we exemplify a way to quantify the performance of the self-calibration for a generic spectro-photometric survey using \emph{mock-up tests} with randomly generated sky catalogues.
The mock-up relative response function $\resp$ used is representative of a plausible variation due to the telescope optics.
The inference of the reconstruction function $\reco$ is performed using a Legendre polynomial basis.
A set of metrics has been defined to quantify the goodness (or badness) of the reconstructed $\rreco$ against the mock-up response function $\resp$.
The goodness of the reconstruction is studied against the average number of sources, exposures, number of degrees of freedom, and the degree of the reconstruction basis.

\subsection{Mock-up self-calibration setup}
\label{sec:mockup}

In each mock-up self-calibration test, 500 different synthetic calibration surveys are randomly produced.
Each of the 500 calibration surveys shares the same average number of sources in the field of view and the same number of exposures;
the sources location $(\xi,\eta)_k$, their magnitude, the exposure sky coordinates $(\xi,\eta)_i$, and their orientation angles are randomly extracted in each realization.
An illustration of a mock-up self calibration survey is displayed in figure~\ref{fig:1-resp-focalplane}-\emph{left}.

In a given realization, the synthetic sky catalogue is created with a fixed number density of sources in random locations of the sky, extracted uniformly in the range $\xi \in (-3,3)$ and  $\eta \in (-3,3)$. The exposure sky coordinates are uniformly extracted in a narrower central region: $\xi \in (-1,1)$ and  $\eta \in (-1,1)$.

In order to obtain a realistic distribution of count rates in our simulations, we take the distribution of stellar magnitudes from the Besan\c{c}on synthetic model of the Galaxy~\citep{besanconmodel}, approximating it with a power law fit over a suitable magnitude range of interest. 
We arbitrarily considered stars with AB magnitudes between \mablow\ and \mabhigh, assuming that stars brighter than $\mab = \mabhigh$ would saturate the detector pixels and stars fainter than $\mab = \mablow$ are too low in signal-to-noise ratio to be reliably used for calibration purposes.

For a reference Galactic latitude of $80 ^\circ$, the number of calibration bright stars of a given $\mab$ expected in the field of view scales as:
\begin{equation}
\label{eq:generalnumbermag}
N(\mab) \propto   \, 10^{0.26 \, (\mab -12)}.
\end{equation}
The star magnitudes are then converted to intrinsic detector  count rates by assuming a constant spectral energy density, the wavelength integration range, and an ideal quantum efficiency over this range.
In the case of a slitless spectroscopic survey, in order to increase the signal-to-noise ratio and avoid detector saturation due to bright sources, the source counts can be obtained by integrating the first order spectrum over a proper wavelength range (e.g. $500 \, \mathring{A}$ in~\citep{Markovic2017}).
A more accurate modelling of the source population and their respective spectrum is beyond the scope of this study.

The conversion from star magnitudes to counts on the detector was computed starting from previous simulations. Sources with magnitude $\mab = \mablow$ correspond to count rates $r_{\rm low}$ about $10^4 / (\expvisnisps)$; bright sources with magnitude $\mab = \mabhigh$ correspond to count rates $r_{\rm high}$ about $10^6 / (\expvisnisps)$.

The relative response function $\resp$ used in the mock-up tests is decreasing toward the edges of the focal plane, by a few percents, and has radial symmetry at a first approximation.
We parametrize the mock-up relative response function as a superimposition of a quadratic polynomial and a few sinusoidal terms.
The radially-symmetric behaviour is representative of vignetting; the sinusoidal terms have been included to further complicate the response and stress the reconstruction procedure. 

We employed three levels of complexity on top of the mock-up response function.
In the first level, we assume a single unsegmented detector with a uniform gain over the whole focal plane; in the second level, the detector is segmented in four equal parts separated by gaps of about $5\%$ of the field of view side; in the third complexity level, the four detector sectors have different relative gains in the range $100\%-90\%$.

The response functions used the mock-up tests is shown in figure~\ref{fig:1-resp-focalplane}; all the mock-up tests of complexity level 1 use the function in figure~\ref{fig:1-resp-focalplane}-\emph{left}; the mock-up test for complexity level 2 uses the same function, with the addition of the inter detector gaps; for complexity level 3, the function in figure~\ref{fig:1-resp-focalplane}-\emph{right} is used. In the following figures, the mock-up response function is referred to as \emph{Power2}+\emph{Fourier2}.

In the mock-up self-calibration tests, the response function $\resp$ and the reconstruction function $\rreco$ are not parametrized in the same form (\emph{e.g.} with the same basis).
This mimics a realistic situation, in which the instrumental response is unknown.
Following the results of the validation tests (\ref{sec:validation}), we use the Legendre polynomials as reconstruction basis.

\subsection{Metrics for the goodness of reconstruction}

\label{sec:metrics}
The metrics to quantify the goodness of the reconstructed $\rreco$ are:
the maximum absolute difference, the cumulative absolute difference, and the unusable fractions.

The \emph{maximum absolute difference} (MAD) is defined as the maximum of the absolute difference between the mock-up $\resp$ and the reconstructed $\rreco$,
\begin{equation}
\mad \, := \, \mbox{max} \left( \, \left |  \resp  - \rreco \right | \right) \, ,
\end{equation}
evaluated on the whole domain of the focal plane excluding the gaps between the detector sectors.

The \emph{cumulative absolute difference} (CAD) is defined as the spatial integral of the absolute difference between the mock up $\resp$ and the reconstructed $\resp$:
\begin{equation}
\cad \, := \, \frac{\int_{\rm{FP}} \, \left |  \resp  - \rreco \right | \, {\rm d}S \,}
{\int_{\rm{FP}} \, {\rm d}S },
\end{equation}
where the surface integral runs over the whole focal plane excluding the gaps between the detector sectors.

The \emph{unusable fraction}, given a \emph{threshold} (UF(th$\%$)), is defined as the fraction of the focal plane
where the absolute deviation $|  \resp  - \rreco | $ exceeds a certain threshold.
In particular, we use the unusable fraction for absolute deviations above $0.7\%$, i.e. $\ufoseven$.

\subsection{Results of mock-up tests}
\label{sec:results}

This section reports the relevant results obtained in the mock-up tests.

In all our mock-up tests, the iterative minimization procedure (\ref{sec:iterative}) converged.
In the mock-up tests with one single unsegmented detector, the minimum of $\kisqe$ is found within a few tens of iterations. The case with four detectors separated by gaps, each one with the same gain, does not present any substantial difference. 
In the mock-up tests with four detector sectors with different gains, the number of iterations needed is about a few hundreds.

As expected, the mock-up tests show that increasing both the average number of sources and the number of exposures is likely to improve the goodness of the reconstruction.
The $\kisqe$ is good if the reconstruction basis degree is sufficiently high.
The focal plane can then be reconstructed (and thus calibrated) to high accuracy.

We studied the trends of the worst (maximum) $\mad$ and $\cad$ values among the 500 realizations of each scenario.
Figure~\ref{fig:worst-MAD-CAD-leg6} shows the values of the $\mad$ and $\cad$ metrics, against the number of sources in the field of view and the number of exposures, restricted to the realizations with the worst values.
The basis of the reconstructed $\reco$ is a Legendre polynomial basis with maximum degree 6 in $x$ or $y$ (for maximum degree 6 we mean a basis with $N=28$ terms, including all the polynomials up to a total power of 6).

The plots show that increasing both the average number of sources and the number of exposures is likely to improve the goodness of the reconstruction. The trends also suggest that the number of degrees of freedom $\ndf$ (equation~\ref{eq:ndf}) of the realization is a driver of the goodness of reconstruction:
realizations with similar number of observations (average number of sources times exposures, at a crude approximation) have similar values of $\mad$ or $\cad$ when the number of observations is low, and tend towards an asymptotic value of $\mad$ or $\cad$ when the number of observations is high.

Figure~\ref{fig:tot-CAD-4567} and figure~\ref{fig:tot-UF07-4567} show the scatter plots of the $\cad$ and $\ufoseven$ metrics respectively,
against the number of degrees of freedom of the realization.
The basis of the reconstructed $\reco$ is a Legendre polynomial basis with maximum degree 4 and 6.

The scatter plots show how the number of degrees of freedom drives the goodness of reconstruction.
The values of the metrics in realizations with the same number of degrees of freedom are contained within a band, which becomes narrower by increasing the number of degrees of freedom. The lower side of the band (good reconstruction) reaches an asymptotic value, which is strongly driven by the maximum degree of the reconstruction basis:
only by increasing the degree of the basis, the asymptotic value can jump to lower values, allowing better reconstructions.
Similar results are found for the $\mad$ and the other UF metrics.

In order to easily compare the trends, we compute the median and the $10\%-90\%$ quantiles of each of the metric distributions.
We show in figure~\ref{fig:trend} the trends of the median and quantiles of each metric against the number of degrees of freedom and the maximum degree of the Legendre basis,
for the $\mad$, $\cad$, and $\ufoseven$ metrics respectively. 
The trends are shown both for the scenario with one single unsegmented detector and in the case with four detectors with small gaps between them and a different relative gain $\gs$ in each detector sector.

The trend plots show that it is possible to reach a good level of calibration even with the additional complexity introduced by the estimation of the $\{ \gs \}$'s in $\rreco$, as long as $\ndf$ is sufficiently high.

In our mock-up tests, with the mock-up response function in figure~\ref{fig:1-resp-focalplane}, working with a number of degrees of freedom about 1000 allows one to obtain a calibration maximum discrepancy of less than $0.7\%$ in a fraction of the focal plane larger than $99\%$ (see figure~\ref{fig:trend}-\emph{bottom}).
In our simulations, $\ndf \approx 1000$ can be reached either with a mean number of 60 sources in the field of view and 20 exposures, or with 30 sources and at least 30 exposures.

Figure~\ref{fig:mock-residuals-1sector} and \ref{fig:mock-residuals} show the reconstructed $\rreco$ and the residuals $(\resp - \rreco) / \delta \rreco$ in two different mock-up tests with $\ndf \sim 4000$ and a reconstruction with a Legendre polynomial basis with maximum degree 6 and 8 respectively. 
The mock-up test displayed in figure~\ref{fig:mock-residuals-1sector} is performed with a single detector sector and using the input response function $\resp$ represented in figure~\ref{fig:1-resp-focalplane}-\emph{left}.
The reconstructed $\reco$ (figure~\ref{fig:mock-residuals-1sector}-\emph{left}) approximates the input $\resp$ (figure~\ref{fig:1-resp-focalplane}-\emph{left}) with high accuracy. The residuals of $\reco$ in every point of the focal plane are contained between $+3.3$ and $-3.5$, showing that the uncertainty is also properly estimated (figure~\ref{fig:mock-residuals-1sector}-\emph{right}).
The mock-up test displayed in figure~\ref{fig:mock-residuals} is performed with 4 detector sectors and using the input response function $\resp$ represented in figure~\ref{fig:1-resp-focalplane}-\emph{right}.
Similarly, the reconstructed $\rreco$ (figure~\ref{fig:mock-residuals}-\emph{left}) accurately approximates the input $\resp$ (figure~\ref{fig:1-resp-focalplane}-\emph{right}) and the residuals are contained between $+3$ and $-3$ (figure~\ref{fig:mock-residuals}-\emph{right}).

Similar studies can be performed for specific surveys and instruments to infer quantitative information about the self-calibration survey layout needed to reach a target accuracy.
The precise number of calibration sources and exposures needed to reach a target accuracy is affected by the complexity of the instrument response function, the detector noise level, and the distribution of the sources.
Nevertheless, the trends shown in this work could represent a plausible scenario.

\section{Conclusions}
This work illustrates and quantifies a technique for the in-flight relative flux self-calibration method, which generalizes the procedure outlined in~\citep{holmes2012designing}.
The technique can be applied for the in-flight calibration of a generic spectro-photometric instrument.

The method is based on the repeated observations of sources in different positions of the focal plane, following a random observation pattern.
The procedure is based on a $\ki$ statistical inference where the reconstruction function accounts for a smooth continuum variation, due to telescope optics, on top of a discontinuous effect due to the segmentation of the detector in different sectors.
The method provides an unbiased inference of the count rates of the sources and of the reconstructed relative response function, in the limit of high count rates.

Mock-up tests have been used to study the convergence of the reconstructed function to an arbitrarily complicated instrument response. We show that the reconstruction also works in case where the detector is segmented in macro-sectors separated by small gaps.

We show that in this procedure the number of repeated observations drives the goodness of the reconstruction. This means that a small number of exposures can be compensated by a large number of sources in the field of view, or vice-versa.
If the number of repeated observations is sufficiently high, it is possible to reconstruct the relative instrument response function with high accuracy, without any prior knowledge.

This work can help defining the self-calibration plan for future large scale surveys, and is particularly useful for space missions whose observation time is subject to tight schedules. 

\ack
Authors are grateful to the Euclid Consortium and in particular Y.~Copin and the whole OU-SIR group, P.~Schneider, G.~Zamorani, M.~Sauvage and K.~Jahnke for the useful discussions including possible future applications of our method to the NISP instrument of the Euclid experiment, and for the help in the review of the document.

Simulations and computations in this work have been performed at the computing facilities of INFN, Sezione di Genova: authors wish to thank the INFN LSF personnel in Genova for their precious and constant support.

\appendix

\section{The iterative minimization}
\label{sec:iterative}
This appendix details the algorithm used to minimize the $\kisqe$~(\ref{eq:kisq}) and solve the system~(\ref{eq:kisqminim}).

\subsection{Partial derivatives with respect to the source rates}

In the partial derivative of the $\kisqe$ with respect to a source rate, the rate term is linear.
The intrinsic source rate $\rate$ that minimizes the $\kisqe$, while keeping all the parameters of $\rreco$ fixed, is given by
\begin{equation}
\label{eq:dkisqdrarrow}
\dkisqdr =  0 \quad \longrightarrow \quad 
\rate =  \frac{\sumik \, \cnt t_i  \frac{ \rrecofpk}{\sig} } { \sumik \, t^2_i \frac{ \rrecofpk^2}{\sig}} \, .
\end{equation}

\subsection{Partial derivatives with respect to the response coefficients}

The partial derivatives of the $\kisqe$ with respect to the $\{ \ql \}$, with $\ell \neq 0$,
can be computed  explicitly:
\begin{eqnarray}
\label{eq:dkisqdq}
\dkisqdq &= -2 \sumk \rate\ \sumik  \frac{t_i \basis \fpxyk} {\sig}  \sum_s \gs \Thetas \\
 &\times  \left [\cnt - \rrecofpk \rate t_i \right] \nonumber \, .
\end{eqnarray}
The condition $\dkisqdq = 0$ can be written in a convenient form by equating~(\ref{eq:dkisqdq}) to zero, then expanding $\rreco$.
In the following, we indicate with $\wlik$ the basis term $\basis \fpxyk$, and with $g_{s(i,k)}$ the gain $g_s$ in the sector $s$ at focal plane coordinates $\fpxyk$.
In the expansion of the continuous function we conveniently use the letter $m$ for the coefficient sum index,
instead of the usual $\ell$ index, which is reserved for the derivative variable $\ql$ in~(\ref{eq:dkisqdq});
also, we separate the $m = 0$ term in the sum, and define
\begin{equation}
\reco = q_0 w_0  \, + \, \sum_{m=1}^N q_m w_m (x,y) \, ,
\end{equation}
then, we rearrange the terms:
\begin{equation}
\label{eq:dkisqdqeqzero1}
\eqalign{
\sum_{m=1}^N  q_m \sumk \rate^2 \sumik \frac{ \gsik^2 t_i^2}{\sig} \wlik \wmik
= \cr
= \sumk \rate \sumik \frac{\gsik t_i \wlik}{\sig} \left [\cnt - \gsik q_0 w_0 \rate t_i  \right].}
\end{equation}
The right side term in equation~(\ref{eq:dkisqdqeqzero1}) involves a \emph{linear combination} of the response coefficients $\qvec$, excluding $q_0$;
the left side term of~(\ref{eq:dkisqdqeqzero1}) is a constant term, which depends on $\ell$.
The set of equations~(\ref{eq:dkisqdqeqzero1}) evaluated for $\ell = 1 \dots N$ represents then an inhomogeneous linear system in the coefficients $\qvec$ 
(from $q_1$ to $q_N$).
The linear system can be represented in a compact matrix form
by defining the vector of constants $\Dvec$  (the right term of equation~\ref{eq:dkisqdqeqzero1}), whose components are
\begin{eqnarray}
\label{eq:Deltal}
\Delta_\ell &:=  \sumk \rate \sumik \frac{\gsik t_i \basis \fpxyk}{\sig} \\
&\times \left [\cnt - \gsik q_0 w_0 \rate t_i  \right] \, . \nonumber
\end{eqnarray}
and the linear system matrix $\tens{H}^{\rm(q)}$, whose entries $\Hlm$ are
\begin{eqnarray}
\Hlm &:=
\sumk \rate^2 \\
&\times \sumik \frac{\gsik^2 t_i^2}{\sig} \basis \fpxyk w_m \fpxyk \,. \nonumber
\end{eqnarray}
The linear system~(\ref{eq:dkisqdqeqzero1}) simply becomes
\begin{equation}
\label{eq:dkisqdqeqzero2}
\tens{H}^{\rm(q)} \, \qvec = \Dvec \,,
\end{equation}
and if the matrix $\tens{H}^{\rm(q)}$ is non-singular, the system can be solved.

In summary, the response coefficients $\{ \ql \}$'s ($\ell = 1, \dots, N)$ which minimize the $\kisqe$, while keeping all the source rates $\{ \rate \}$'s and relative gains $\{ \gs \}$'s fixed,
are obtained by solving the linear system given by the~(\ref{eq:dkisqdq}) equated to zero:
\begin{equation}
\label{eq:dkisqdqarrow}
\dkisqdq =  0 \quad (\ell = 1, \dots, N) \, \longrightarrow \quad \ql = {H^{\rm(q)}}^{-1}_{\ell m} \Delta_m \, .
\end{equation}

The response coefficient $q_0$ is treated differently with respect to the other $\{ \ql \}$'s, in order to cure the degeneracy between $\rate$ and the scale of $\resp$ outlined in section~\ref{sec:obs}.
The coefficient $q_0$ is the one that multiplies the basis $w_0$, which is uniform across the focal plane
(e.g. it is equal to 1 for the power basis and the Legendre basis, and to $\frac 1 4$ for the Fourier basis);
thus, the term $q_0 w_0$ in the expansion of the continuous response function $\reco$
is a constant term which determines a global shift in the values returned by the function,
and in particular, it adjusts the value in the focal plane origin $\hat f(0,0 \, | \qvec)$.

Once all the $\{ \ql \}$'s with $\ell \neq 0$ are estimated, the normalization~(\ref{eq:normalizationql}) becomes a constraint on $q_0$:
\begin{equation}
\label{eq:constraintq0}
q_0 = \frac 1 {w_0} \left[ 1- \sum_{\ell=1}^N \ql \basis (0,0) \right].
\end{equation}
The constrain is used to derive the value of $q_0$.

It is worth mentioning a very special solution of the system~(\ref{eq:dkisqdqeqzero2}), the one of the \emph{ideal} detector 
i.e. a detector with no statistical fluctuations and with a uniform relative response on the whole focal plane;
in this special ideal case, $\cnt  = \rate t_i$. In the polynomial basis $q_0 = w_0^{-1}$ and all the other $\qvec$'s are zero;
following the definition in~(\ref{eq:Deltal}), in the ideal case, $\Dvec$ is a \emph{null} vector, and the solution of the system~(\ref{eq:dkisqdqeqzero2})
is that the components of $\qvec$ (from 1 to $N$) are zero: the response is uniform on the whole focal plane.

\subsection{Partial derivatives with respect to the gains}
In the partial derivative of the $\kisqe$ with respect to a gain, the gain term is linear.
The relative gain $\gs$ in the detector sector $s$ that minimizes the $\kisqe$, while keeping all the source rates $\{ \rate \}$'s and the coefficients $\{ \ql \}$'s fixed, is given by
\begin{equation}
\label{eq:dkisqdgarrow}
\dkisqdg =  0 \quad \longrightarrow \quad 
\gs = \frac { \sumk \rate \sumiks \cnt  t_i \frac{ \recofpk }{\sig}} { \sumk \rate^2 \sumiks t_i^2 \frac{ \recofpk^2 }{\sig}} 
\end{equation}
Only the $\nks$ observations of the source $k$ where the focal plane coordinates $\fpxyk$ are in the sector $s$ are included in the sum.
The gain $g_c$ of the sector where the origin of the focal plane coordinate is located is always fixed to one.

\section{Statistical uncertainties}
\label{sec:uncertainties}
\newcommand{\ratep}{r_{k^\prime}}
\newcommand{\gsp}{g_{s^\prime}}
\newcommand{\ddkisqdrdr}{\frac{\partial^2 \kisqe}{\partial \rate \partial \ratep}}
\newcommand{\ddkisqdqdq}{\frac{\partial^2 \kisqe}{\partial \ql \partial q_m}}
\newcommand{\ddkisqdgdg}{\frac{\partial^2 \kisqe}{\partial \gs \partial \gsp}}
\newcommand{\ddkisqdrdq}{\frac{\partial^2 \kisqe}{\partial \rate \partial \ql}}
\newcommand{\ddkisqdrdg}{\frac{\partial^2 \kisqe}{\partial \rate \partial \gs}}
\newcommand{\ddkisqdqdg}{\frac{\partial^2 \kisqe}{\partial \ql \partial \gs}}
\newcommand{\Hrr}{H_{k k^\prime}^{\rm(rr)}}
\newcommand{\Hrq}{H_{k \ell}^{\rm(rq)}}
\newcommand{\Hqq}{H_{\ell m}^{\rm(qq)}}
\newcommand{\Hgg}{H_{s s^\prime}^{\rm(gg)}}
\newcommand{\Hgq}{H_{s \ell}^{\rm(gq)}}
\newcommand{\Hqg}{H_{\ell s}^{\rm(qg)}}
\newcommand{\Hrg}{H_{k s}^{\rm(rg)}}
\newcommand{\Hgr}{H_{s k}^{\rm(gr)}}
\newcommand{\covrr}{\mbox{Cov}[\rate, r_{k^\prime}]}
\newcommand{\covqq}{\mbox{Cov}[\ql, q_m]}
\newcommand{\covgg}{\mbox{Cov}[\gs, \gsp]}
\newcommand{\covrq}{\mbox{Cov}[\rate, q_\ell]}
\newcommand{\covqr}{\mbox{Cov}[q_\ell, \rate]}
\newcommand{\covrg}{\mbox{Cov}[\rate, \gs]}
\newcommand{\covgr}{\mbox{Cov}[\gs, \rate]}
\newcommand{\covgq}{\mbox{Cov}[\gs, \ql]}
\newcommand{\covqg}{\mbox{Cov}[\ql, gs]}

The \emph{statistical uncertainties} of the intrinsic source rates $\{ \drate \}$,
of the response coefficients  $\{ \dql \}$, and of the relative gains $\{ \dgs \}$ can be estimated from the diagonal elements
of the covariance matrix, computed as the inverse of the (halved) second derivatives matrix 
of the $\kisqe$ (the \emph{Hessian} matrix).

The second partial derivatives of the $\kisqe$ with respect to the $k$-th and the $k^\prime$-th source intrinsic rates are
\begin{equation}
\ddkisqdrdr \, = \, 2 \delta_{k k^\prime} \sumik \frac{ \rrecofpk^2 t_i^2}{\sig} \, .
\end{equation}

The second partial derivatives of the $\kisqe$ with respect to the  $\ell$-th and the $m$-th relative response coefficients, where $\ell$ and $m$ are not zero, are
\begin{eqnarray}
\ddkisqdqdq \, &= \, 2 \sumk \rate^2  \\
&\times \sumik \frac{ \gsik^2 t_i^2}{\sig} \basis \fpxyk w_m \fpxyk \, . \nonumber
\end{eqnarray}

The second partial derivatives of the $\kisqe$ with respect to the $s$-th and the $s^\prime$-th sector relative gain are
\begin{equation}
\ddkisqdgdg \, = \, 2 \delta_{s s^\prime} \sumk \rate^2 \sumiks \frac{ \recofpk^2 t_i^2}{\sig} \, .
\end{equation}

The second partial derivatives of the $\kisqe$ with respect to the $k$-th source intrinsic rate and the $\ell$-th relative response coefficient are
\begin{eqnarray}
\label{eq:Hdef}
\ddkisqdrdq \, &=  \, 2 \sumik \frac{ \gsik \basis \fpxyk t_i} {\sig} \\
 &\times \left [2 \rrecofpk \rate t_i - \cnt \right ] \, . \nonumber
\end{eqnarray}

The second partial derivatives of the $\kisqe$ with respect to the $k$-th source intrinsic rate and the $s$-th relative sector gain are
\begin{eqnarray}
\label{eq:Hdef}
\ddkisqdrdg \, &=  \, 2 \sumk \sumiks \frac{\recofpk t_i} {\sig} \\
 &\times \left [2 \rrecofp \rate t_i - \cnt \right ] \, . \nonumber
\end{eqnarray}

The second partial derivatives of the $\kisqe$ with respect to the $\ell$-th relative response coefficient and the $s$-th relative sector gain are
\begin{eqnarray}
\label{eq:Hdef}
\ddkisqdqdg \, &=  \, 2 \sumk \rate \sumiks \frac{\basis \fpxyk t_i} {\sig} \\
 &\times \left [2 \rrecofp \rate t_i - \cnt \right ] \, . \nonumber
\end{eqnarray}
In the partial derivatives with respect to the gain $\gs$, the sum over the exposures ($i$ label) includes only the observations where the focal plane coordinates are in the sector $s$.

We define the matrices of the (halved) second partial derivatives, $\Hrr$, $\Hqq$, $\Hgg$, $\Hrq$, $\Hrg$, and $\Hqg$ whose elements are
\begin{equation}
\Hrr \, := \, \frac 1 2 \ddkisqdrdr \, ,
\end{equation}
\begin{equation}
\Hqq \, := \, \frac 1 2 \ddkisqdqdq \, , 
\end{equation}
\begin{equation}
\Hgg \, := \, \frac 1 2 \ddkisqdgdg \, , 
\end{equation}
\begin{equation}
\Hrq \, := \, \frac 1 2 \ddkisqdrdq \, ,
\end{equation}
\begin{equation}
\Hrg \, := \, \frac 1 2 \ddkisqdrdg \, ,
\end{equation}
\begin{equation}
\Hqg \, := \, \frac 1 2 \ddkisqdqdg \, ,
\end{equation}
and the full second derivatives matrix $\tens{H}$:
\begin{equation}
\label{eq:H}
\tens{H} :=
\left(
\begin{array}{c|c|c}
\Hrr & \Hrq & \Hrg\\
\hline
\Hqr & \Hqq & \Hqg\\
\hline
\Hgr & \Hgq & \Hgg
\end{array}
\right),
\end{equation}
The matrices $\Hrr$ and $\Hgg$ are diagonal, and the matrix $\Hqq$ is symmetric. The entries corresponding to the partial derivatives with respect $q_0$ are excluded from $\tens H$; we explain in the next paragraphs how to compute variance and covariances of $q_0$. The entries corresponding to the partial derivatives with respect to the gain of the central sector $g_c$ are also excluded from $\tens H$, since they are a row (or column) of zeroes, and would make $\tens H$ singular.

The variance-covariance matrix $\tens{C}$ is the inverse of the matrix $\tens{H}$.
The variance-covariance matrix can be represented as a block matrix:
\begin{equation}
\label{eq:Cov}
\tens{C} =
\left(
\begin{array}{c|c|c}
\covrr & \covrq & \covrg\\
\hline
\covqr & \covqq & \covqg\\
\hline
\covgr & \covgq & \covgg
\end{array}
\right) \, .
\end{equation}
The diagonal elements of the variance-covariance matrix $\tens{C}$ are the squares of the marginalized $1\sigma$ statistical uncertainties on the corresponding parameters:
the set of diagonal entries of the top-left block of $\tens C$ are the squares of the uncertainties on the intrinsic source rates $\{ \drate \}$;
the following diagonal entries are the squares of the uncertainties on the relative response coefficients $\{ \dql \}$ (excluding $\ell = 0$); the sets of diagonal entries of the bottom-right block of $\tens C$ are the squares of the uncertainties on the relative gains $\{ \dgs \}$, excluding the uncertainty on the central sector gain $\delta g_c$, which is identically zero.

The variance of $q_0$ and the covariance between $q_0$ and the other $\{ \ql \}$'s can be directly obtained from the constraint~(\ref{eq:constraintq0}),
interpreted as a linear composition of the correlated random variables $\{ \ql \}$'s, with coefficients $\basis(0,0)/w_0(0,0)$.
The variance of $q_0$ is:
\begin{eqnarray}
\label{eq:varq0}
\mbox{Var}[q_0] &= \langle q_0^2 \rangle - \langle q_0 \rangle^2 \\
&= \frac 1 {w_0^2}\sum_{\ell=1}^N \sum_{m=1}^N \basis (0,0) w_m (0,0) \, \mbox{Cov}[\ql, q_m] \nonumber
\end{eqnarray}
and its square root is interpreted as the marginalized $1\sigma$ statistical uncertainty $\delta q_0$.
In matrix form, eq.~\ref{eq:varq0} reads:
\begin{equation}
\mbox{Var}[q_0] = \frac 1 {w_0^2} \, \wvect(0,0) \, \covqqvec \wvec(0,0) \, ,
\end{equation}
where $\covqqvec$ is the matrix corresponding to the $\covqq$ block in $\tens{C}$ and $\wvec(0,0)$ is the vector containing the basis elements (with $\ell \neq 0$) evaluated in the focal plane origin.

The covariance between $q_0$ and another $\ql$ is:
\begin{eqnarray}
\label{eq:covq0ql}
\mbox{Cov}[q_0, \ql]  &=  \langle q_0 \ql \rangle - \langle q_0 \rangle \langle \ql \rangle \\ 
&= - \frac 1 {w_0}\sum_{\ell^\prime=1}^N  w_{\ell^\prime} (0,0) \, \mbox{Cov}[q_{\ell^\prime}, \ql] \, .\nonumber
\end{eqnarray}
In matrix form, eq.~\ref{eq:covq0ql} reads:
\begin{equation}
    \mbox{Cov}[q_0, \ql] = - \frac 1 {w_0} \wvect(0,0) \, \covqqvec \, .
\end{equation}
In the case of the polynomial basis (\ref{sec:power}), the variance $\mbox{Var}[q_0]$, as well as the covariances $\mbox{Cov}[q_0, \ql]$, are identically zero,
since the only non-vanishing $\basis(0,0)$ is $w_0$, which equals 1.

The complete variance-covariance matrix $\tens{C}_{\rm q}$ of the relative response parameters $\{ \ql \}$ (with $\ell = 0, \dots, N)$ is therefore
\begin{equation}
\label{eq:Covq}
\tens{C}_{\rm q} =
\left(
\begin{array}{c|c}
\mbox{Var}[q_0] & \mbox{Cov}[q_0, \ql] \\
\hline
\mbox{Cov}[\ql, q_0] & \covqq
\end{array}
\right) \, .
\end{equation}

The variance-covariance matrix $\tens{C}_{\rm q}$ is used to compute the statistical uncertainty
of the continuous part of the reconstructed response function $\reco$, evaluated at a given point of the focal plane with coordinates $(x,y)$.
The variance of $\reco$ can be directly obtained from the expansion~(\ref{eq:expansion1}),
interpreted as a linear composition of the correlated random variables $\{ \ql \}$'s, with coefficients $\basis(x,y)$:
\begin{eqnarray}
\label{eq:dfq}
\mbox{Var}[\reco)] &= \langle \reco ^2 \rangle - \langle \reco \rangle^2 \\
&= 
\sum_{\ell=0}^N \sum_{m=0}^N \basis (x,y) w_m (x,y) \, \mbox{Cov}[\ql, q_m] \nonumber
\end{eqnarray}
In matrix form, eq.~\ref{eq:dfq} reads:
\begin{equation}
    \mbox{Var}[\reco)] = \wvect(x,y) \, \tens{C}_{\rm{q}} \, \wvec(x,y) \, .
\end{equation}

The square root of $\mbox{Var}[\reco)]$ is interpreted as the marginalized $1\sigma$ statistical uncertainty
on the relative reconstructed response function $\reco$, at focal plane coordinates $(x,y)$.

The statistical uncertainty on the reconstructed response function $\rreco$ evaluated at coordinates $(x,y)$ can be estimated from the expansion~(\ref{eq:expansion2}), interpreted as a composition of the correlated random variables $\{ \ql \}$'s and $\{ \gs \}$'s:
\begin{eqnarray}
\label{eq:dfqg}
\mbox{Var}&[\rreco)] = \langle \rreco ^2 \rangle - \langle \rreco \rangle^2 \nonumber \\ 
&= \,
(g^2 + \delta g^2)\sum_{\ell=0}^N \sum_{m=0}^N \basis (x,y) w_m (x,y) \mbox{Cov}[\ql, q_m] \nonumber \\
&+ \, \delta g^2 \, \reco^2 \nonumber \\
&+ \, 2 g \reco \sum_{\ell=1}^N \basis (x,y) \mbox{Cov}[\ql, g] \nonumber \\
&+ \, \left( \sum_{\ell=1}^N \basis (x,y) \mbox{Cov}[\ql, g] \right)^2 \, ,
\end{eqnarray}
where $g$ is the gain in the sector where the coordinate $(x,y)$ belongs to, $\delta g$ is its variance, and $\mbox{Cov}[\ql, g]$ is the vector of covariances between the coefficient $\ql$ and the gain $g$. The covariance $\mbox{Cov}[q_0, g]$ is identically zero.
In matrix form, eq.~\ref{eq:dfqg} reads:
\begin{eqnarray}
\mbox{Var}&[\rreco)] = (g^2 + \delta g^2) \wvect(x,y) \, \vec{C}_{\rm{q}} \, \wvec(x,y)\nonumber\\
&+ \, \delta g^2 \, \reco^2 \nonumber \\
&+ \, 2 g \, \reco  \, \wvect(x,y) \, \covqgvec \nonumber \\
&+ \, \left( \wvect(x,y) \, \covqgvec \right)^2 \, ,
\end{eqnarray}
where $\covqgvec$ is the vector representation of $\mbox{Cov}[\ql, g]$.

For the central sector, $g=1$, $\delta g=0$, and the covariance vector $\mbox{Cov}[\ql, g]$ is null, therefore equation~(\ref{eq:dfqg}) simplifies to equation~(\ref{eq:dfq}). This simplification also holds in the case of an unsegmented  detector, where only one sector exists.

The square root of $\mbox{Var}[\rreco)]$ is interpreted as the marginalized $1\sigma$ statistical uncertainty
on the relative reconstructed response function $\rreco$, at focal plane coordinates $(x,y)$.

\section{Validation of the procedure}
\label{sec:validation}

\begin{figure*}[tpb]
\centering
\includegraphics[width=.5\textwidth]{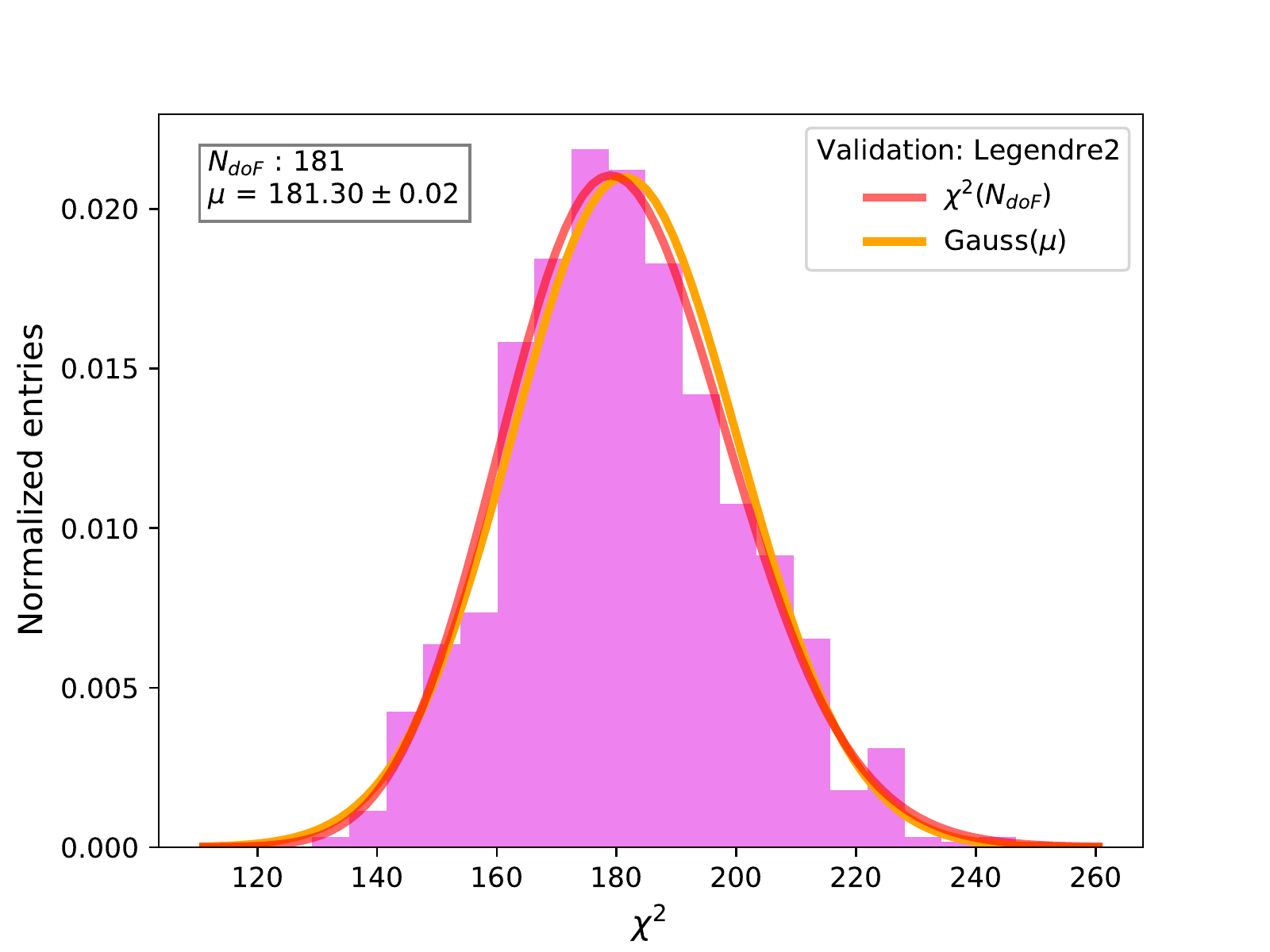}\hfill
\includegraphics[width=.5\textwidth]{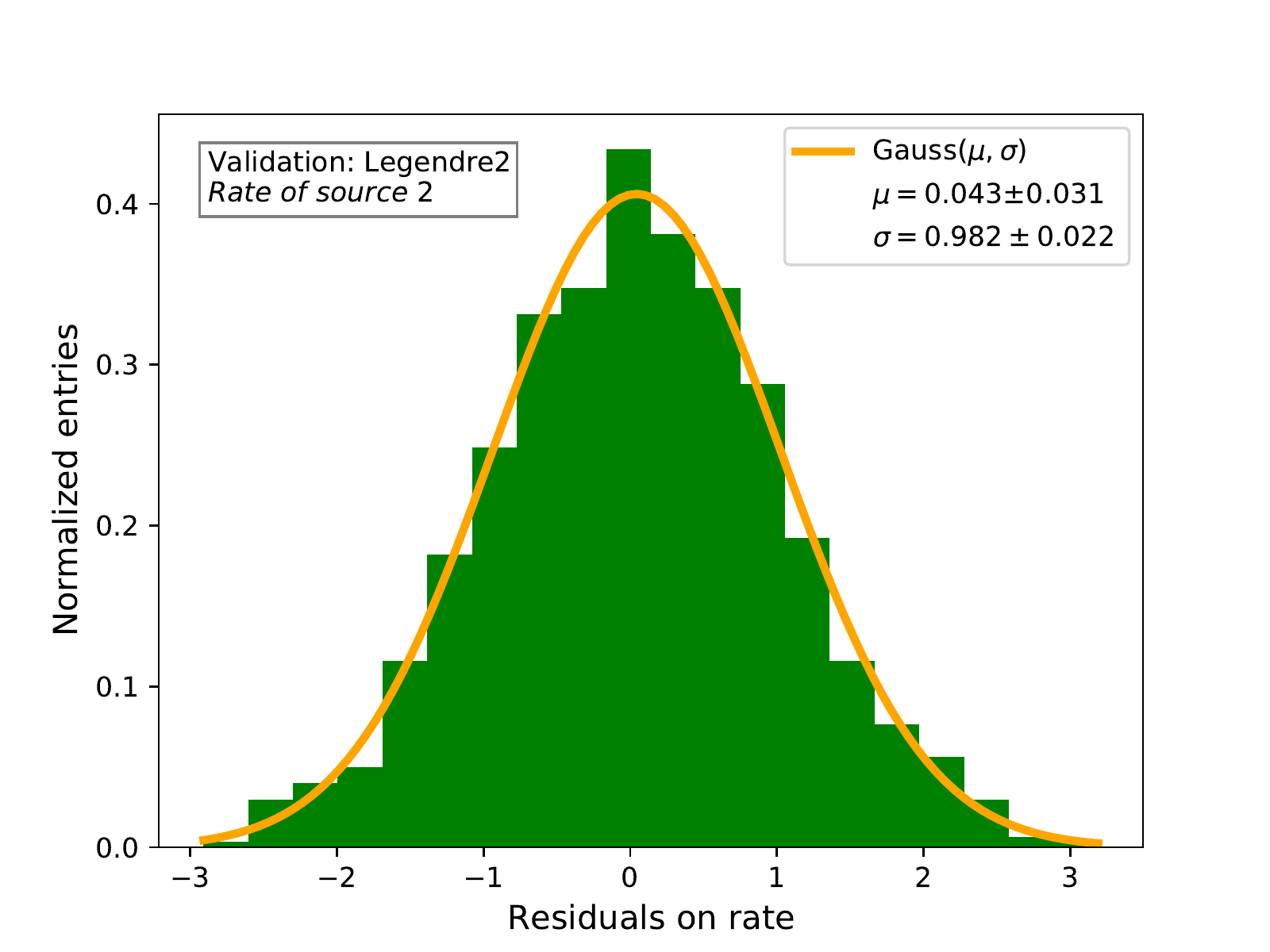}

\caption{Validation of the inference procedure: $\kisqe$ and $\{ \rate \}$'s.
\emph{Left-} The resulting $\kisqe$ distribution in a validation test with $\ndf=181$;
a theoretical $\kisq$ distribution with $\ndf$ degrees of freedom is overlaid to the distribution (red line);
a Gaussian fit to the distribution is also overlaid (orange line).
\emph{Right-} A rate residual $(\rate^{\rm{true}} - \rate)/ \delta \rate $ distribution in a validation test;
a Gaussian fit to the distribution is overlaid (orange line).
}
\label{fig:validation1}
\end{figure*}

\begin{figure*}[tpb]
\centering
\includegraphics[width=.5\textwidth]{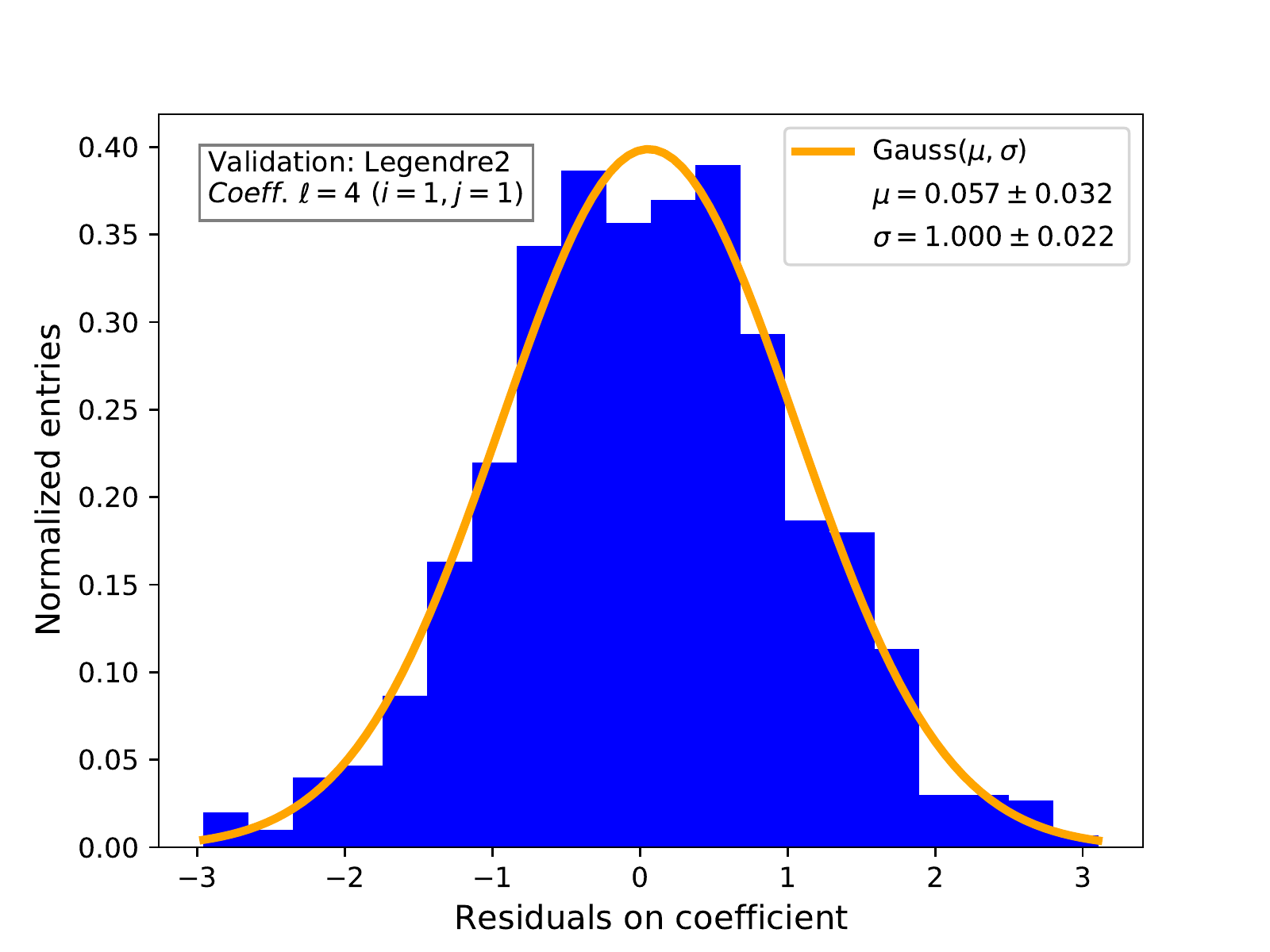}\hfill
\includegraphics[width=.5\textwidth]{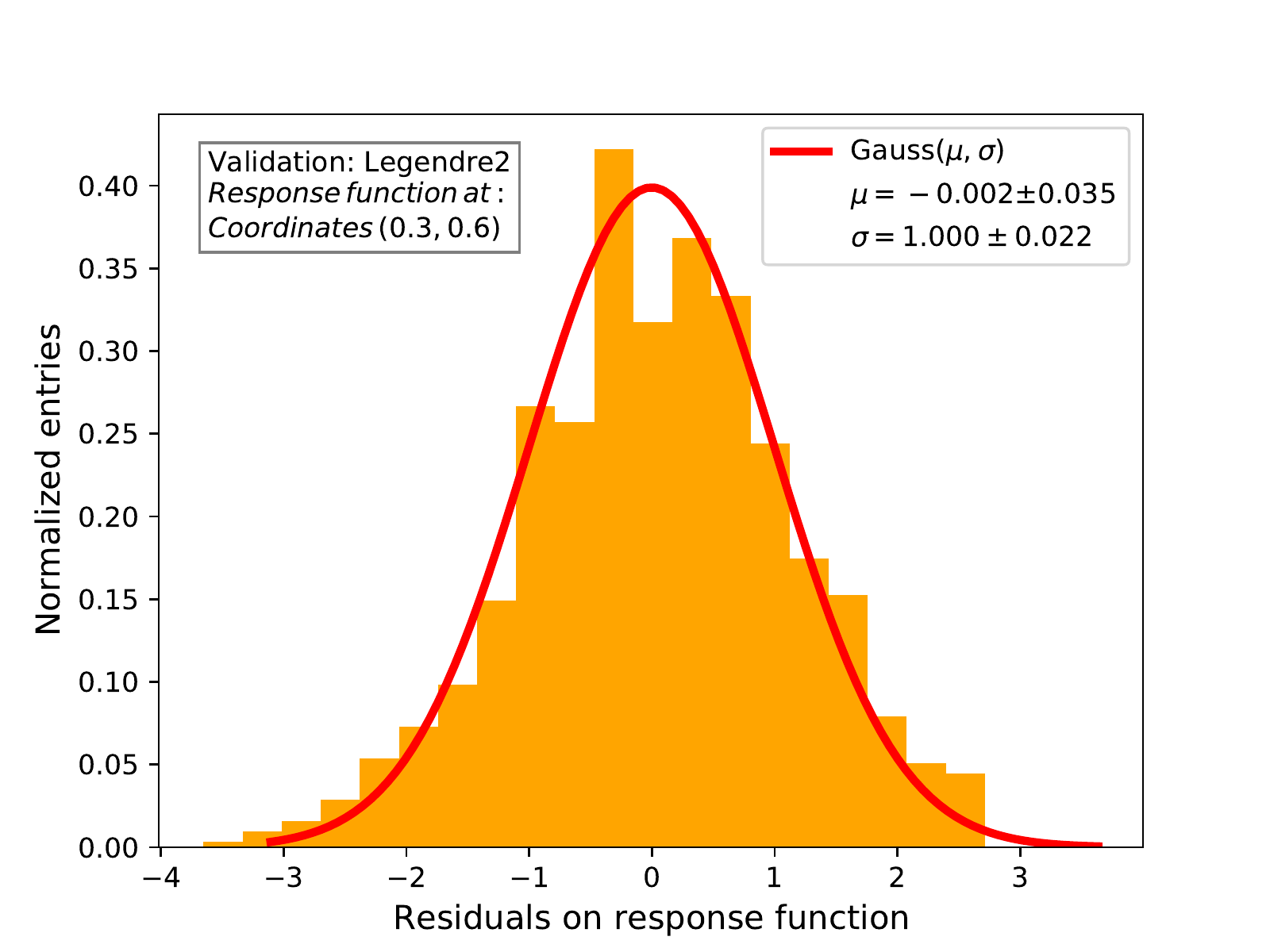}

\caption{Validation of the inference procedure: $\{ \ql \}$'s and $\rreco$.
\emph{Left-} A coefficient residual $(\ql^{\rm{true}} - \ql)/ \delta \ql $ distribution in a validation test, using a Legendre basis, for $\ell = 4$;
a Gaussian fit to the distribution is overlaid (orange line).
\emph{Right-} Residual distribution $(f^{\rm{true}} - \hat f)/ \delta \hat f $  of the parametric reconstructed relative response function $\reco$ 
at focal plane coordinates $(0.3, 0.6)$  in a validation test, using a Legendre basis;
a Gaussian fit to the distribution is overlaid (red line).
}
\label{fig:validation2}
\end{figure*}

Validation tests are set up by producing synthetic calibration surveys with known intrinsic source rates $\{ \rate \}$-\mbox{true} and using a relative response function $\resp$ of the same form of the reconstruction response function $\reco$, with known response coefficients $\{ \ql \}$-\mbox{true} and relative gains $\{ \gs \}$-\mbox{true}.
For a given configuration of the sky catalogue, the exposures, and the response function, 990 synthetic calibration surveys are produced, each survey using a different random seed for the extraction of the observed count (section~\ref{sec:obs});
the 990 realizations of the synthetic calibration sets $\{\cnt, \sig, \fpxy, t_i\}$ therefore differ in the detector counts $\cnt$ and its (population) variance $\sig$, but have the same sets of focal plane coordinates $\fpxy$ and exposure times $t_i$.
In a given validation configuration, synthetic sky catalogues (section~\ref{sec:sky}) are created with 20 sources in random position of the sky $(\eta, \xi)$ extracted uniformly within $-1$ and $+1$.
Each source rate is such to give $\cnt$ between $10^4$ and $10^6$; the detector noise is about $10^3$ counts;
16 synthetic exposures (section~\ref{sec:exposures}) are created by extracting uniformly the sky coordinates pointing (within $-1$ and $+1$)  and the orientation angle.

In the validation tests, the relative response function $\resp$ is parametrized using the same form of the reconstruction function $\reco$:
for example a power set (or Fourier, or Legendre) basis with $N$ coefficients is used as $\resp$,
and then the reconstruction $\reco$ is parametrized with the same power set (or Fourier, or Legendre) basis with $N$ coefficients.
The coefficients $\{ \ql \}$-\mbox{true} are chosen in the percent range, usually with signs that provide a decreasing $\resp$ towards the focal plane edges.
The constant-basis coefficient $q_0$ is determined such that $\resp$ returns one in the center of the focal plane.

The minimization procedure (section~\ref{sec:kisqminim}) is performed for each of the 990 realizations.
The minima of the $\kisqe$ are compared against the expected $\ndf$ and the inferred intrinsic source rates $\{ \rate\}$'s and the relative response coefficients $\{ \ql \}$'s are compared against the configured $\{ \rate \}$-\mbox{true} and $\{ \ql \}$-\mbox{true}.

The minima of the $\kisqe$ returned by the iterative minimization procedure are indeed distributed as a $\kisq$ distribution with $\ndf$ degrees of freedom.
Figure~\ref{fig:validation1}-\emph{left} shows the resulting $\kisqe$ distribution in a validation test.
The resulting distribution is well described by a $\kisq$ distribution with $\ndf$ degrees of freedom (red line).
A Gaussian fit to the distribution (orange line) returns a mean value coherent with the asymptotic behaviour of the $\kisq$ distribution,
that for large $\ndf$ tends to a Gaussian with $\ndf$ as mean, and twice $\ndf$ as variance.
The minimum value of the $\kisqe$ returned by the iterative minimization procedure can be then used as indicator of the goodness of the model.

The validation tests show that the residuals for the intrinsic source rates $\{ \rate\}$'s are unbiased:
all the residuals $ (\rate^{\rm{true}} - \rate)/ \delta \rate $ produced are well described by a Gaussian  with mean compatible with zero and variance compatible with one.
Figure~\ref{fig:validation1}-\emph{right} shows an example of a residual distribution for the intrinsic source rate $r_2$ in a validation test.

Each inferred coefficient $\{ \ql \pm \delta \ql \}$ of the relative response $\reco$ is compared against the configured $\{ \ql \}^{\rm{true}}$.
The residual $ (\ql^{\rm{true}} - \ql)/ \delta \ql $ for each $\ell$ is computed for each of the 990 realizations.
The validation procedure confirms that the resulting residual distributions are unbiased:
all the $\{ \ql \}$ residual distributions produced are well described by a Gaussian  with mean compatible with zero and variance compatible with one.
Figure~\ref{fig:validation2}-\emph{left} shows an example of a residual distribution for the coefficient $q_4$ of the Legendre basis.

Finally, the validation tests confirm that the parametric reconstructed relative response function $\rreco$ is unbiased.
The residual $ (f^{\rm{true}} - \hat f)/ \delta f $ at a given focal plane coordinate $(x,y)$ is computed for each of the 990 realizations.
The validation procedure confirms that the resulting residual distributions are unbiased:
all the $\rreco$ residual distributions produced at various coordinates $(x,y)$ are well described by a Gaussian  with mean compatible with zero and variance compatible with one.
Figure~\ref{fig:validation2}-\emph{right} shows an example of a residual distribution for $\rreco$ evaluated at coordinates $(0.3, 0.6)$, using the Legendre polynomial basis.

Validation tests have been repeated with different numbers of sources, different numbers of exposures, and using different response bases and numbers of coefficients.
In all the configurations tested, the results were unbiased.
Our studies showed that the non-central blocks in the second derivatives matrix must be included to produce unbiased results, especially if the number of sources or exposures is low.

\section{Expansion of the reconstruction response function}
\label{sec:expansion}

The continuous reconstruction function $\reco$ is expanded as:
\begin{equation}
\label{eq:expansion1}
\reco = \sum_{\ell=0}^N \ql \, \basis (x,y) \, = \sum_{i=0}^n \sum_{j=0}^n p_{i,j} v_i (x) v_j (y) \, .
 \end{equation}
This appendix details the mapping between the pair $(i,j)$ and $\ell$ and provides examples of reconstruction bases.

\subsection{Ordering and mapping of the two- and one- dimensional sets}
\label{sec:ordering}
Any given pair of indices $(i,j)$ is one-to-one mapped into its corresponding $\ell$ index in the expansion of equation~(\ref{eq:expansion1}).
The mapping convention we use is the following:
\begin{itemize}
\item The $\ell=0$ term  corresponds to $(i,j) = (0,0)$.
\item The  $\ell=1$ and $\ell=2$ terms respectively correspond to $(1,0)$ and $(0,1)$.
\item In the successive set terms of $\ell$,
the rank of the $x$ basis $v_i(x)$ is first raised to the lowest un-expanded term (e.g. $i=2$ this time),
with the rank of the $y$ basis $v_j(y)$ set to zero ($j=0$).
\item Each successive $\ell$ in the set is found by decreasing the $i$ index and increasing the $j$ index of one unit,
until the $\ell$ corresponding to $i=0$ and $j$ raised to the lowest un-expanded term (e.g. $j=2$ this time).
\item The procedure is then repeated with the next lowest one-dimensional un-expanded term.
\end{itemize}

The first terms of the expansion for the different bases are reported in the next section.

\subsection{Reconstruction bases}
\label{sec:power}
\newcommand{\pni}{p^{(n)}_i}
\newcommand{\cpni}{ \{ \pni \}}

The reconstruction function can be expanded using any two dimensional basis in a closed interval.
In this work, we studied the response with the set of power basis, the Legendre polynomial basis, and the Fourier basis.

The set of \emph{powers} $\{ t^i \}$, e.g. $\{1, \, t, \, t^2,  \, t^3 , \dots \}$, is a basis for functions $f(t)$ of one real variable $t$ defined in a closed interval.
The convergence of a sequence of polynomials is described in detail in textbooks of mathematical methods (see for example chapter 5.4 of~\citep{Byron1992}). 

The \emph{two-dimensional power basis} set $\{ \basis (x,y) \}$ on the focal plane is constructed from multiplications of powers $\{ x^i \}$ and $\{ y^j \}$, with the ordering delineated in~\ref{sec:ordering}.
In particular, the first few $\{ \basis (x,y) \}$ of the power basis are $\{ 1,  \, x, \, y,  \, x^2,  \, xy,  \, y^2, \, x^3, \, x^2y, \, xy^2, \, y^3, \, x^4, \, x^3y, \, \dots \}$.

The power basis has some advantages deriving from the normalization constraint~(\ref{eq:normalizationql}), which simplifies to $q_0 = 1$, since the only non-vanishing $\basis(0,0)$ is $w_0$, which equals one. Also, the variance $\mbox{Var}[q_0]$, as well as the covariances $\mbox{Cov}[q_0, \ql]$, are identically zero.

\label{sec:legendre}
The set of the \emph{Legendre polynomials} $\{ P_i(t) \}$ forms a complete \emph{orthogonal} basis over the closed interval $t \in [-1,1]$.
The first few Legendre polynomials are: $\{1, \, t, \, \frac 1 2 (3t^2 -1), \, \frac 1 2 (5 t^3 -3t), \, \dots \}$.

The \emph{two-dimensional Legendre basis} set $\{ \basis (x,y) \}$ on the focal plane
is constructed from multiplications of Legendre polynomial $\{ P_i(x) \}$ and $\{ P_j(y) \}$,
with the ordering delineated in~\ref{sec:ordering}.
The first few $\{ \basis (x,y) \}$ of the Legendre basis are
$\{ 1,  \, x, \, y,  \, \frac 1 2 (3x^2 -1),  \, xy,  \, \frac 1 2 (3y^2 -1), \, \frac 1 2 (5x^3 - 3x), \, \frac 1 2 (3 x^2 -1)y,  \, \dots \}$.
The two-dimensional basis $\{ \basis (x,y) \}$ defined above is  orthogonal.

The Legendre polynomial basis has some advantages deriving from the completeness. As expected, we find out that the iterative minimization procedure needs fewer iterations to converge with the Legendre polynomial basis. 

A square-integrable function on a closed interval can be expanded with the \emph{Fourier series}, where the set of trigonometric functions forms an orthonormal basis. 
The convergence of the Fourier series is treated in many textbooks of mathematical methods (for example see theorems $5.4$ and $5.5$ in~\citep{Byron1992}).

We conveniently reorder the one-dimensional Fourier basis $\{ v_i (t) \}$
as:\\ $\{ \frac 1 2, \, \sin(\pi t), \, \cos(\pi t), \, \sin(2\pi t), \, \cos(2\pi t), \, \dots \}$.
The \emph{two-dimensional Fourier basis} set $\{ \basis (x,y) \}$ on the focal plane
is constructed from multiplications of the Fourier basis $\{ v_i(x) \}$ and $\{ v_j(y) \}$,
with the ordering delineated in~\ref{sec:ordering}.
The first few $\{ \basis (x,y) \}$ of the Fourier basis are\\
$\{ \frac 1 4,  \,  \frac 1 2  \sin( \pi x), \, \frac 1 2 \sin(\pi y), \, 
\frac 1 2 \cos(\pi x),  \, \sin( \pi x) \sin (\pi y), \\
\frac 1 2 \cos (\pi y), \, \frac 1 2 \sin( 2 \pi x), \,  \cos(\pi x) \sin(\pi y),  \, \dots \}$.


\bibliography{bibliography}

@article{holmes2012designing,
author = {Holmes, Rory and Hogg, David W and Rix, Hans-Walter},
journal = {Publications of the Astronomical Society of the Pacific},
number = {921},
pages = {1219},
publisher = {IOP Publishing},
title = {{Designing imaging surveys for a retrospective relative photometric calibration}},
volume = {124},
year = {2012}
}

@article{redbook,
archivePrefix = {arXiv},
arxivId = {astro-ph.CO/1110.3193},
author = {Laureijs, R and Amiaux, J and Arduini, S and Augueres, J-L and Brinchmann, J and Cole, R and Cropper, M and Dabin, C and Duvet, L and Ealet, A},
eprint = {1110.3193},
journal = {arXiv preprint arXiv:1110.3193},
primaryClass = {astro-ph.CO},
title = {{Euclid Redbook}},
year = {2011}
}

@misc{besanconmodel,
    title    = "Besan\c{c}on model of stellar population synthesis of the Galaxy",
    url = {https://model.obs-besancon.fr},
    year = {2019}
}

@misc{roman,
    title    = "Roman Space Telescope/NASA: mission overview",
    url = {https://roman.gsfc.nasa.gov},
    year = {2021}
}

@book{Byron1992,
author = {Byron, Frederick W. and Fuller, Robert W.},
title = {{Mathematics of Classical and Quantum Physics}},
ISBN = "048667164X",
year = {1992}
}

@article{Shafer2015,
    author = {Shafer, Daniel L. and Huterer, Dragan},
    doi = {10.1093/mnras/stu2640},
    issn = {13652966},
    journal = {Monthly Notices of the Royal Astronomical Society},
    month = {mar},
    number = {3},
    pages = {2961--2969},
    publisher = {Oxford University Press},
    title = {Multiplicative errors in the galaxy power spectrum: Self-calibration of unknown photometric systematics for precision cosmology},
    volume = {447},
    year = {2015}
}

@article{Padmanabhan2008,
archivePrefix = {arXiv},
arxivId = {astro-ph/0703454},
author = {Padmanabhan, Nikhil and Schlegel, David J. and Finkbeiner, Douglas P. and Barentine, J. C. and Blanton, Michael R. and Brewington, Howard J. and Gunn, James E. and Harvanek, Michael and Hogg, David W. and Ivezi{\'{c}}, {\v{Z}}eljko and Johnston, David and Kent, Stephen M. and Kleinman, S. J. and Knapp, Gillian R. and Krzesinski, Jurek and Long, Dan and {Neilsen, Jr.}, Eric H. and Nitta, Atsuko and Loomis, Craig and Lupton, Robert H. and Roweis, Sam and Snedden, Stephanie A. and Strauss, Michael A. and Tucker, Douglas L.},
doi = {10.1086/524677},
eprint = {0703454},
issn = {0004-637X},
journal = {The Astrophysical Journal},
month = {feb},
number = {2},
pages = {1217--1233},
primaryClass = {astro-ph},
publisher = {American Astronomical Society},
title = {{An Improved Photometric Calibration of the Sloan Digital Sky Survey Imaging Data}},
url = {http://www.sdss.org},
volume = {674},
year = {2008}
}

@article{Markovic2017,
archivePrefix = {arXiv},
arxivId = {1606.07061},
author = {Markovi{\v{c}}, Katarina and Percival, Will J. and Scodeggio, Marco and Ealet, Anne and Wachter, Stefanie and Garilli, Bianca and Guzzo, Luigi and Scaramella, Roberto and Maiorano, Elisabetta and Amiaux, J{\'{e}}r{\^{o}}me},
doi = {10.1093/mnras/stx283},
eprint = {1606.07061},
issn = {13652966},
journal = {Monthly Notices of the Royal Astronomical Society},
month = {jun},
number = {3},
pages = {3677--3698},
publisher = {Oxford University Press},
title = {{Large-scale retrospective relative spectrophotometric self-calibration in space}},
volume = {467},
year = {2017}
}

@techreport{Schlafly2012,
archivePrefix = {arXiv},
arxivId = {1201.2208v2},
author = {Schlafly, E F and Finkbeiner, D P and Juric, M and Magnier, E A and Burgett, W S and Chambers, K C and Grav, T and Hodapp, K W and Kaiser, N and Kudritzki, R.-P and Martin, N F and Morgan, J S and Price, P A and Rix, H.-W and Stubbs, C W and Tonry, J L and Wainscoat, R J},
eprint = {1201.2208v2},
title = {{PHOTOMETRIC CALIBRATION OF THE FIRST 1.5 YEARS OF THE PAN-STARRS1 SURVEY}},
year = {2012}
}

@article{Amendola2018,
archivePrefix = {arXiv},
arxivId = {1606.00180},
author = {Amendola, Luca and Appleby, Stephen and Avgoustidis, Anastasios a J. and Percival, Will and Pettorino, Valeria and Porciani, Cristiano and Quercellini, Claudia and Read, Justin and Rinaldi, Massimiliano and Sapone, Domenico and Sawicki, Ignacy and Scaramella, Roberto and Skordis, Constantinos and Simpson, Fergus and Taylor, Andy and Thomas, Shaun and Trotta, Roberto and Verde, Licia and Vernizzi, Filippo and Vollmer, Adrian and Wang, Yun and Weller, Jochen and Zlosnik, Tom},
journal = {Living Reviews in Relativity},
doi = {10.1007/s41114-017-0010-3},
eprint = {1606.00180},
issn = {14338351},
month = {dec},
number = {1},
pages = {67},
publisher = {Springer},
title = {{Cosmology and fundamental physics with the Euclid satellite}},
url = {https://doi.org/10.1007/s41114-017-0010-3},
volume = {21},
year = {2018}
}


\end{document}